\documentclass{article}

\usepackage[english]{babel}

\usepackage[letterpaper,top=2cm,bottom=2cm,left=3cm,right=3cm,marginparwidth=1.75cm]{geometry}

\usepackage{algorithm}
\usepackage{algpseudocode}
\usepackage{amsmath, amsfonts, amssymb}
\usepackage[normalem]{ulem}
\usepackage{graphicx}
\usepackage[colorlinks=true, allcolors=blue]{hyperref}
\usepackage{authblk}
\usepackage{lineno}
\usepackage{enumitem}
\usepackage[
  backend      = biber,
]{biblatex}
\setlist{
    topsep=1pt,
    partopsep=0pt,
    itemsep=0pt,
    parsep=1pt
}

\newcommand\td{\text{d}}

\newcommand\RB{Rayleigh-B\'enard }

\newcommand\mup{\mu_m}
\newcommand\sigp{\sigma_M}
\newcommand\rp{r_{i, M}}
\newcommand\muo{\mu_\mathrm{obs}}
\newcommand\sigo{\sigma_\mathrm{obs}}
\newcommand\ro{r_{i,\mathrm{obs}}}

\newcommand\MA{\mathcal{A}}
\newcommand\MB{\mathcal{B}}

\newcommand\MN{\mathcal{N}}
\newcommand\MO{\mathcal{O}}
\newcommand\MV{\mathcal{V}}
\newcommand\MW{\mathcal{W}}

\newcommand\EE{\mathbb{E}}
\newcommand\BE{\mathbb{E}}
\newcommand\BR{\mathbb{R}}
\newcommand\R{\BR}

\title{Probabilistic data-driven turbulence closure modelling by assimilating statistics}

\author[1]{Sagy R. Ephrati\footnote{(sagy@chalmers.se)}}
\affil[1]{Department of Mathematical Sciences, Chalmers University of Technology and University of Gothenburg, 412 96 Gothenburg, Sweden}

\date{}

\begin{document}
\maketitle
\begin{abstract}
A framework for deriving probabilistic data-driven closure models is proposed for coarse-grained numerical simulations of turbulence in statistically stationary state. The approach unites the \textit{ideal large-eddy simulation} model \cite{langford1999optimal} and data assimilation methods. The method requires \textit{a posteriori} measured data to define a stochastic large-eddy simulation model, which is combined with a Bayesian statistical correction enforcing user-specified statistics extracted from high-fidelity flow snapshots. Thus, it enables computationally cheap ensemble simulations by combining knowledge of the local integration error and knowledge of desired flow statistics. A model example is given for two-dimensional \RB convection at Rayleigh number $\mathit{Ra}=10^{10}$, incorporating stochastic perturbations and an ensemble Kalman filtering step in a non-intrusive way. Physical flow dynamics are obtained, whilst kinetic energy spectra and heat flux are accurately reproduced in long-time ensemble forecasts on coarse grids. The model is shown to produce accurate results with as few as 20 high-fidelity flow snapshots as input data. 
\end{abstract}

\section{Introduction} \label{sec:introduction}
The highly turbulent nature of fluid flows provides a major challenge in the prediction of fluid-dynamical processes.
Examples include oceanic and atmospheric flows, relevant to geophysics and climate science, or engineering and industrial applications involving thermal processes. 
The nonlinearity in the governing equations causes interaction between flow features of different scales, leading to energy distribution over a wide range of scales of motion \cite{pope2001turbulent}. 
Accurate direct numerical simulations (DNS) of fluid-dynamical models thus require very fine computational grids to fully resolve all flow features, and quickly become prohibitively expensive to carry out. 
Feasible simulation strategies for such systems therefore necessitate a reduction in computational complexity to approximate the flow evolution. 
Well-established approaches dealing with complexity reduction include reduced-order modelling \cite{snyder2022reduced}, where the governing equations or associated operators are formulated in a low-dimensional manner, and large-eddy simulation (LES) \cite{sagaut2005large, geurts2022direct}, where the governing equations are spatially filtered and small-scale effects are modelled. 
In the current work, we focus on the LES approach and combine this with stochastic forecasting and data assimilation.
The result is a general modelling framework for data-driven error correction in low-fidelity fluid predictions, which we demonstrate for two-dimensional \RB convection.
We elaborate on the various model components below.

\paragraph{Large-eddy simulation}
The spatially filtered equations of motion form the starting point for large-scale LES models. 
The underlying rationale is that computationally feasible methods can be developed by only capturing the largest scales of motion, since these are most important for the overall flow evolution. 
The level of detail in the LES solution is determined by the filter width. 
Explicit LES follows from applying the adopted spatial filter to the governing equations directly. 
Alternatively, the equations can be filtered implicitly by discretizing differential operators on coarse computational grids. 
Regardless of the chosen approach, filtering introduces errors and uncertainty in the flow evolution. 
The filter generally does not commute with the nonlinear terms in the governing equations, yielding a so-called commutation error. 
In addition, discretization error is introduced on coarse computational grids due to poorly resolved spatial derivatives. 
One can add modelling terms to the governing equations with the aim of compensating for the errors introduced by filtering and coarsening \cite{geurts2002alpha, piomelli2015grid, rouhi2016dynamic}. 
We will refer to this as the LES model, although it is also often referred to as the closure model or the sub-filter or sub-grid scale model. 
Furthermore, filtering introduces uncertainty since it discards information of small spatial scales. 
Infinitely many fully resolved flow fields will correspond to the same filtered LES field, which makes it impossible to know the exact evolution of the LES solution. 
Consequently, it is desirable to develop error-correcting LES models that simultaneously quantify the uncertainty in the LES solution.

The abstract ideal LES model was derived by Langford and Moser \cite{langford1999optimal} and can serve as a guideline for model development. The model is ideal in the sense that it reproduces single-time, multi-point statistics of the unfiltered solution and minimizes error in the instantaneous dynamics. It states that the ideal evolution of an LES solution is the average evolution conditional to an infinite number of fully resolved (unfiltered) fields corresponding to the current large-scale (filtered) field. One can probe this distribution of fields using a stochastic forecast ensemble and subsequently approximate the conditional average. Naturally, the model depends on the governing equations, the adopted filter and, in simulations, on the chosen spatial and temporal discretizations. This makes it challenging to approximate the conditional average well. However, empirical approximations can still be found through the use of data.

The field of data-driven LES has seen rapid development in recent years due to the increasing availability of computational resources and accessibility of high-fidelity data. Machine learning is used regularly to train neural networks as LES models \cite{sanderse2024scientific, beck2019deep}. One can distinguish between \textit{a priori} and \textit{a posteriori} learning, which differ in their objective function. \textit{A priori} methods rely only on the high-fidelity data to learn a closure model, for example by minimizing a function of the unfiltered and the filtered solution. This approach can yield accurate short-time predictions, but might also suffer from errors and instabilities due to model-data inconsistency \cite{duraisamy2021perspectives, agdestein2024discretize}. \textit{A posteriori} methods can overcome this inconsistency, by directly comparing the results of the low-fidelity solver and the high-fidelity data \cite{frezat2022posteriori}. Consequently, the model will depend on the configuration of the low-fidelity solver. The presently proposed model uses an \textit{a posteriori} approach based on the ideal LES formalism to calibrate a stochastic subgrid-scale model.

\paragraph{Stochastic modelling in coarsened fluid descriptions}
Stochasticity is commonly used in geophysical fluid dynamics to model uncertainty \cite{hasselmann1976stochastic, buizza1999stochastic} stemming from imperfect initial conditions and incomplete models \cite{palmer2019stochastic}. A prevalent interpretation of stochastic forcing is that it models the chaotic behavior of unresolved small scales on the resolved dynamics. For example, data-informed stochastic models have been developed for the two-scale Lorenz '96 system \cite{lorenz1996predictability}. This system serves as a low-dimensional model of the atmosphere with variables evolving over different time scales. By replacing the influence of fast variables on the slow variables with data-driven stochastic processes, \cite{arnold2013stochastic} were able to account for model error. The addition of time-correlated noise further improved the forecasting skill. Other approaches based on data-driven Markov chains \cite{crommelin2008subgrid} have led to state-dependent stochastic processes modelling the effects of unresolved variables yielding a good reproduction of statistics of the resolved variables, such as probability density functions and temporal autocorrelation. Alternatively, \cite{boral2023neural} produce an ensemble of predictions in latent space via data-driven neural stochastic differential equations and subsequently average the corresponding LES predictions to approximate the ideal LES prediction. They report reduced errors over long simulation times and good predictions of the time-averaged energy spectrum when compared to implicit LES and deterministic neural network closures.

The current method fits in the recent modelling trend in observational sciences where high resolution simulations are replaced by stochastically forced simulations on coarse grids. 
New types of data-driven stochastic models have been developed for fluid-dynamical systems for the purpose of modelling uncertainty in observations and correct for undesired coarsening effects. 
For example, the effect of transport noise \cite{holm2015variational, memin2014fluid} in advection-dominated geophysical flows has been studied in the context of uncertainty quantification \cite{cotter2019numerically, resseguier2020data, ephrati2023data} and data assimilation \cite{cotter2020particle}. 
These studies showed good quantification of uncertainty in advective processes at severely reduced computational costs. 
Another advantage of stochastic forcing is that it allows for indefinite simulation of signals with desired statistical properties, for example mimicking features of the dataset from which the model is calibrated.
The chaotic evolution of turbulent flows prompts the development of stochastic models that reproduce flow statistics rather than pointwise agreement with some reference. 
Inspired by the nonlocality of turbulence, a space- and time-dependent model can be decomposed via global basis functions for which only the time series are modeled \cite{frederiksen2024statistical, frederiksen2017stochastic}, possibly based on statistical data. 
Examples include the use of include proper orthogonal functions (POD) \cite{berkooz1993proper} and Fourier modes to reproduce their respective spectra in coarse numerical simulations of \RB convection \cite{ephrati2023rbpod, ephrati2023rb}. \cite{frederiksen2006dynamical} employed spherical harmonic functions to develop a stochastic subgrid-scale term with memory effects for barotropic flow, and found good agreement in the resulting energy spectra.
Capturing flow statistics instead of the full flow also reduces the amount of data required to calibrate the model. 
Low-dimensional models with basis functions tailored to user-specified statistics can be derived \cite{edeling2020reducing} and used to develop stochastic models at severely reduced computational costs while accurately reproducing selected statistics, as was recently shown for the two-dimensional Navier-Stokes equations \cite{hoekstra2024reduced}.

\paragraph{Data assimilation and error correction}
Data assimilation (DA) deals with efficiently incorporating data in predictions and is widely employed in geophysical sciences. 
DA combines predictions of dynamical systems with observations optimally, taking into account the uncertainty in both these aspects \cite{law2015data, reich2015probabilistic}. 
A common way to include information into a prediction ensemble is via Bayes' theorem \cite{rosic2013parameter, reich2015probabilistic}. 
This has led to DA schemes for nonlinear problems such as the ensemble Kalman filter (EnKF) \cite{evensen1994sequential, evensen2003ensemble} and particle filters \cite{bain2009fundamentals}.
In the context of DA, filtering\footnote{The term \textit{filtering} is used both the fields of LES and data assimilation but carries a different meaning in each of the fields. 
In LES, filtering refers to spatial filtering: a method to determine the level of physical detail to keep in the LES solution. 
In data assimilation, filtering deals with the sequential updating of a probability distribution function of a random variable as new measurements become available.} of turbulent systems has also been carried out by representing interactions of resolved and unresolved scales as stochastic processes \cite{majda2012filtering}. 
Recent work has provided a theoretical framework for filtering of dynamical systems by observing statistics only \cite{bach2024filtering}, with correcting model errors as a possible application. 
The observations are often dealt with sequentially in DA, resulting in `on-the-fly' updates of predictions or parameters as new information becomes available. 
This contrasts with data-driven LES, where model parameters are commonly determined before performing a numerical simulation. 
Nonetheless, we will show here that DA techniques can be adapted to yield turbulence closure models for coarsened discretized systems.

Coarsening a discretized system will lead to a change in the predicted dynamics. 
Such a change is not necessarily a cause for concern at sufficiently high resolution. 
However, severe coarsening induces strong errors and yields numerical solutions converging to a different statistically steady state, as may be observed when measuring statistics of the dynamical system at large lead times. 
These structural errors can be alleviated by steering predictions towards observations (or specific regions of the state space), thereby increasing their likelihood. 
This is referred to as \textit{nudging} \cite{akyildiz2020nudging}. 
A similar concept of continous data assimilation (CDA) based on interpolated approximations of measurements has been developed for fluid flows \cite{azouani2014continuous}. 
CDA assimilates observations while the predictions are integrated in time via a term that nudges the prediction towards the observation. 
Theoretical convergence for downscaled solutions has been proved for the two-dimensional Navier-Stokes equations \cite{carlson2024super}, accompanied by numerical results for this system \cite{gesho2016computational} and two-dimensional \RB convection \cite{altaf2017downscaling, hammoud2023continuous}. 
The CDA approach is similar to the continuous-time 3D-Var DA scheme \cite{courtier1998ecmwf, blomker2013accuracy}. 
Recently, a heuristic data-driven stochastic closure model was derived from the 3D-Var scheme, with the aim of reproducing reference energy spectra. 
This has been demonstrated for the two-dimensional Euler equations \cite{ephrati2023euler}, \RB convection \cite{ephrati2023rb}, and quasi-geostrophic flow on the sphere \cite{ephrati2023qg}. 
However, unrealistic dynamics may arise when nudging the solution too strongly \cite{reich2015probabilistic} and we therefore endeavour to replace ad-hoc modelling assumptions by a structured method based on DA techniques. 
This leads to a modelling framework from which we derive a nudging approach for turbulent flows. 
One of the results in this paper is that the proposed model yields realistic instantaneous dynamics while reproducing selected flow statistics in long-time simulations, even when only a small amount of data is available.

\paragraph{Contributions and paper outline}
In this paper, we propose a probabilistic data-driven LES closure modelling strategy with the aim of correcting for undesired coarsening effects on the dynamics whilst modelling the inherent uncertainty. 
The approach is inspired by the abstract \textit{ideal LES model} \cite{langford1999optimal} and consists of a stochastic ensemble LES prediction followed by a Bayesian correction.
Specifically, stochastic LES yields a computationally cheap ensemble prediction using \textit{a posteriori} measured data.
Subsequently, a Bayesian correction updates selected key statistics (or quantities of interest, QoIs) based on high-fidelity data.
Thus, we combine knowledge of the local errors of the low-fidelity solver with knowledge of long-time statistics of the high-fidelity reference result. 
The model permits indefinite computationally efficient ensemble simulations of the dynamical system. 
The method is demonstrated for two-dimensional \RB convection, where we focus on kinetic energy spectra and heat flux as key statistics.

The paper is structured as follows. 
In Section \ref{sec:ideal_les}, we recapitulate the ideal LES model and extend this to discrete time to set the stage for stochastic LES modelling.
In Section \ref{sec:model}, we recall sequential data assimilation and formulate the general closure model by assimilating statistics.
The closure model and adopted implementation is described for \RB convection in Section \ref{sec:application}, after which we assess short-time and long-time results in the cases of plenty training data and sparse training data in Section \ref{sec:results}. 
The paper is concluded in Section \ref{sec:conclusions}.

\section{Ideal large-eddy simulation}
\label{sec:ideal_les}
LES generally starts from a filtered description of the governing equations of motion. 
The abstract ideal LES model \cite{langford1999optimal} was derived to compensate for unwanted effects caused by underresolving the turbulent flow. 
It is ideal in the sense that it minimizes the instantaneous error in the evolution of the large-scale dynamics. 
In this section we recall the ideal LES model as presented by Langford and Moser \cite{langford1999optimal}.
We summarize Ideal LES in section \ref{subsec:ideal_les_cont} and subsequently extend the formulation to discrete time in section \ref{subsec:ideal_les_discrete}.
In section \ref{subsec:ideal_les_naive} we motivate stochastic modelling on coarse computational grids using ideal LES.

\paragraph{Notation}
We follow the notation of \cite{langford1999optimal} and denote by $u$ and $v$ unfiltered turbulent fields.
The space of unfiltered fields is denoted by $\MV$. 
We let $w$ be a resolvable large-scale field, referred to as an LES field, and the space of filtered fields is denoted by $\MW$. 
The filtered variables are denoted by a tilde $\tilde{\cdot}$.
Distributions are denoted by $\pi$, which will be distinguished via subscripts.
Time levels are denoted by superscripts, e.g., $u^n$ denotes a turbulent field at time $t^n$.
Sets of fields are written in curly brackets $\{\cdot\}$, where we distinguish between sequences of snapshots and ensembles at specific times.
Sequences of snapshots of a solution $u$ are denoted by $\{u^j\}$.
An ensemble at time $t^n$ is denoted by $\{u_i^n\}$, where the $i^\mathrm{th}$ ensemble member is the field $u_i^n$.
As will become clear from context, we also augment the notation of ensembles with subscripts $f$ and $a$ to respectively denote forecast and analysis ensembles.

\subsection{Ideal LES in continuous time}
\label{subsec:ideal_les_cont}
We consider a field of interest $u$ with an evolution given by $L(u)$, which contains a nonlinear advection term. 
The numerical method that integrates $L$ will be referred to as the high-fidelity solver.
The filtered evolution of $u$ reads \begin{equation}
    \widetilde{\frac{\partial u}{\partial t}} = \widetilde{L(u)},
    \label{eq:filtered_evolution}
\end{equation}
which can be expressed as the evolution of the filtered field $\tilde{u}$ as \begin{equation}
    \frac{\partial \tilde u}{\partial t} = L(\tilde u) + M( u).
    \label{eq:evolution_filtered}
\end{equation}
In the last equation, $M$ is a model term that appears due to the filtering of the nonlinear advection term, which causes a commutation error. Evaluating $M$ requires explicit knowledge of the unfiltered field and will instead be replaced by a model.

The ideal subgrid model is shown to minimize error in the instantaneous evolution of large scales and produce accurate spatial statistics \cite{langford1999optimal}. 
The fundamental insight in the model derivation is that a spatial filter must discard information to be useful. 
In other words, it cannot be invertible and maps from a high-dimensional (possibly infinite-dimensional) space of turbulent fields to a lower-dimensional space of LES fields. 
Therefore, for any LES field $w$, there exists a distribution of turbulent fields $u$ such that $\tilde{u}=w$.
By imposing that any spatial statistic of the true solution is reproduced by the LES solution, it is shown that the ideal LES evolution must satisfy \begin{equation}
    \frac{\td w}{\td t} = \left\langle\widetilde{\frac{\td u }{\td t}} \Bigg| \tilde u = w \right\rangle,
    \label{eq:ideal_LES_evolution}
\end{equation}
where $\langle\cdot|\cdot\rangle$ denotes the conditional average.
In this formulation, it is necessary to think of the turbulent field $u$ as a single realization in an ensemble of turbulent fields.
Thus, equation \eqref{eq:ideal_LES_evolution} describes the average filtered evolution of an ensemble of fields $u$, conditional to the filtered fields $\tilde{u}$ being equal to the LES field $w$.
Equation \eqref{eq:ideal_LES_evolution} generalizes the results of Adrian \cite{adrian1975role}, who showed that modelling turbulence reduces to approximating conditional averages of statistics. 

The ideal LES evolution is rewritten as \begin{equation}
    \frac{\partial w}{\partial t} = L_{\mathrm{LES}}(w) + m(w),
\end{equation}
where $L_{\mathrm{LES}}$ is an approximation of of the right-hand side of the original evolution, for example obtained as a coarse numerical discretization. 
Throughout this paper, we will refer to the numerical method integrating
$L_\mathrm{LES}$ as the low-fidelity solver.
The ideal LES model $m(w)$ is consequently defined as \begin{align}
        m(w) &= \langle M(u) | \tilde{u} = w\rangle, \label{eq:ideal_model} \\
        M(u) &= \widetilde{\frac{\partial u}{\partial t}} - L_\mathrm{LES}(\tilde{u}). \label{eq:ideal_measurement}
\end{align}

The above result holds for general flow configurations, numerical methods, and filters. 
Nonetheless, it is evident that all these choices will influence the model in practical situations. 
The flow domain and physical parameters naturally determine the operator $L$. 
The chosen filter and numerical discretization, which in itself implicitly induces a filter, influence $L_\mathrm{LES}$ and thus also affect the ideal subgrid model. 
However, equations (\ref{eq:ideal_model}-\ref{eq:ideal_measurement}) indicate how to construct the ideal model once knowledge of the full system is available. 
In particular, it suggests that measurements of the subgrid force $M(u)$ in \eqref{eq:ideal_measurement} are suited to serve as input for data-driven models. 
These measurements require integrating the filtered solution in time using the low-fidelity solver, which ensures that effects of spatial and temporal discretization are also measured, and comparing this to the filtered evolution of the unfiltered solution.
As a consequence, the closure model will depend on the adopted numerical method and resolution, which is a necessary feature for consistency of the closure model.

\subsection{Ideal LES in discrete time}
\label{subsec:ideal_les_discrete}
As an intermediate step towards stochastic data-driven LES models, we proceed by expressing ideal LES in discrete time.
This amounts to defining the \textit{ideal distribution} of which the mean is the ideal LES prediction.
This formulation will aid the model description in section \ref{sec:model}.

As before, we treat a turbulent field as a single entity in an ensemble of turbulent fields and adopt a probabilistic notation as follows.
We denote by $\pi_U$ the distribution of unfiltered turbulent fields, where $\pi_U^n(u)$ describes the probability of a turbulent field $u$ being the realization of the flow dynamics at time $t^n$.
We define the high-fidelity forward flow map $\psi_{\Delta t}(u^n) := u^{n+1}=u^n + \int_{t^n}^{t^{n+1}}\! L(u^n)\,\td t$, where the superscripts denote the time levels.
Since the $\psi_{\Delta t}$ is a deterministic map, we can describe the evolution of a distribution of fields as a transition kernel $\tau(v|u)=\delta\left(v-\psi_{\Delta t}(u)\right)$ acting on the distribution $\pi_U^n$ as \begin{equation}
\begin{split}
    \pi_U^{n+1}(v) = \int_\MV\!\delta\left(v-\psi_{\Delta t}(u)\right)\pi_U^n(u)\,\td u,
\end{split}
\end{equation}
where $\delta$ is the Dirac delta function.

Specifically, given an LES solution $w^n$, we define the ideal distribution at $t^{n+1}$ as follows.
We begin with the distribution $\pi_V^n$ of unfiltered fields $u$ that satisfy $\tilde{u}=w^n$.
This distribution evolves according to \begin{equation}
    \pi_V^{n+1}(v) = \int_\MA\! \delta\left(v-\psi_{\Delta t}(u)\right)\pi_V^n(u)\,\td u,
    \label{eq:ideal_distribution}
\end{equation}
where $\MA=\left\{u\in \MV|\tilde{u}=w^n \right\}$.
We use $\pi_V^{n+1}(v)$ to define the distribution $\pi_W^{n+1}(w)$ of filtered fields $w$ that correspond to the unfiltered fields at $t^{n+1}$, \begin{equation}
    \pi_W^{n+1}(w) = \int_\MB\, \pi_V^{n+1}(v)\,\td v,
    \label{eq:filtered_distribution}
\end{equation}
where $\MB=\left\{v\in\MV|\tilde{v}=w\right\}$.
We refer to $\pi_W^{n+1}$ as the ideal distribution at time $t^{n+1}$.
The ideal LES prediction $w^{n+1}$ is then given as the average field in this distribution, \begin{equation}
    w^{n+1} = {\EE_{\pi_W^{n+1}}[w]} = \int_\MW\! w\,\pi_W^{n+1}(w)\,\td w.
    \label{eq:ideal_LES_probabilistic}
\end{equation}

Equations (\ref{eq:ideal_distribution}-\ref{eq:ideal_LES_probabilistic}) describe the following procedure.
Given an LES prediction at time $w^n$, we would like to know the best approximation at time $w^{n+1}$ as described by ideal LES.
We thus consider the distribution $\pi_V^n(u)$ of unfiltered fields $u$ at time $t^n$ that all correspond to the same LES field $w^n$ after applying the filter $\tilde{\cdot}$.
Each unfiltered field in this distribution is evolved in time according to the high-fidelity flow map in \eqref{eq:ideal_distribution}, leading to a distribution $\pi_V^{n+1}(v)$ of unfiltered fields at time $t^{n+1}$.
To find the ideal LES prediction at this time, we have to compute the corresponding distribution $\pi_W^{n+1}(w)$ of filtered fields $w$ as in \eqref{eq:filtered_distribution} and subsequently compute its mean following \eqref{eq:ideal_LES_probabilistic}.

\subsection{A naive Monte Carlo approximation to ideal LES}
\label{subsec:ideal_les_naive}
Equation \eqref{eq:ideal_LES_probabilistic} indicates that the ideal LES prediction $w^{n+1}$ at time $t^{n+1}$ is the mean of the ideal distribution $\pi_W^{n+1}$.
Computing $w^{n+1}$ thus requires the evaluation of the integral on the right-hand side of equation \eqref{eq:ideal_LES_probabilistic}, which is generally intractable since it concerns an infinite number of fluid fields.
Any feasible approximation of this integral will require using Monte Carlo methods, which in this case corresponds to computing the empirical mean of a distribution approximating $\pi_W^{n+1}$.
A naive approximation to the ideal LES prediction $w^{n+1}$ can therefore be obtained by initializing an ensemble $\{u_i^n\}$ with an empirical distribution that approximates $\pi_V^n$, and use this ensemble to subsequently approximate the integrals in equations \eqref{eq:ideal_distribution}, \eqref{eq:filtered_distribution}, and \eqref{eq:ideal_LES_probabilistic}.
This is summarized in algorithm \ref{alg:naive_idealLES}.

Algorithm \ref{alg:naive_idealLES} only serves to illustrate the concept of ideal LES and will not be used for actual prediction of flow fields in the current study.
We refer to this algorithm as `naive' as it quickly becomes computationally intractable.
Since every realization in the Monte Carlo approximation needs to be integrated with the high-fidelity solver, the computational costs grow rapidly even when a modest number of ensemble members are used in the approximation.
To remedy this, we explore in the next section the combination of ideal LES and data assimilation techniques to approximate the ideal distribution without having to integrate at  high spatial resolution.

\begin{algorithm}
    \caption{Naive approximation of ideal LES over one time step}
    \label{alg:naive_idealLES}
    \begin{algorithmic}
        \Procedure{Naive ideal LES}{$w^n$, $L$, $N$, $\tilde{\cdot}$, $\Delta t$}
        \For{$i=1,\ldots,N$}
        \State $u_i^n \gets \mathrm{sample}(\pi_V^n)$
        \Comment{Draw i.i.d. samples from $\pi_V^n$, recall that ${\tilde{u}_i^n}=w^n$}
        \State $v_i^n\gets u_i^n + \int_{t^n}^{t^{n+1}}\! L(u_i^n)\,\td t$
        \Comment{Integrate samples in time to approximate $\pi_V^{n+1}$}
        \State $w_i^n\gets \tilde{v}_i^n$
        \Comment{Filter samples to approximate $\pi_W^{n+1}$}
        \EndFor
        
        \State $w^{n+1}\gets \frac{1}{N}\sum_{i=1}^N w_i^n$
        \Comment{The empirical mean of $\pi_W^{n+1}$ approximates the ideal LES prediction}
        \State \Return $w^{n+1}$
        \EndProcedure
    \end{algorithmic}
\end{algorithm}

\section{Probabilistic closure modelling framework}
\label{sec:model}
The probabilistic description of ideal LES as presented in the previous section motivates using stochastic simulation strategies to obtain empirical approximations to distributions of fields.
Finding probability distributions given observations of the underlying dynamical system has traditionally been the goal of data assimilation \cite{law2015data}, which aims to combine model predictions with observational data in a manner that optimally balances the uncertainties present in both the predictions and the observations.
This is commonly achieved by filtering, i.e., sequentially updating the probability distribution of the variable of interest in two steps referred to as the prediction and the analysis.

In this section, we formulate stochastic LES in the context of sequential data assimilation.
In particular, we discuss applying a Bayesian correction to the stochastic LES prediction in order to approximate the ideal distribution.
By considering flows in statistically stationary states, we relax this modelling criterion and focus on accurate prediction of flow statistics instead. 
Through this combination of ideal LES and data assimilation methods, we arrive at a general framework of closure modelling by assimilating statistics.

Throughout this section, we assume that high-fidelity data is available in the form of flow snapshots.
From these snapshots, we can extract time series and empirical distributions of flow statistics (also referred to as quantities of interest or QoIs) and access this information in the model formulation.

\subsection{Stochastic LES and sequential data assimilation}
\label{subsec:les_sequential_da}

We first briefly recall the ensemble prediction and Bayesian correction steps in sequential data assimilation, which enables us to formulate the general closure model in Section \ref{subsec:general_closure}.

\paragraph{Ensemble LES prediction}
For simplicity, we consider a single LES solution $w^n$ at time $t^n$ and formulate the \textit{prediction ensemble} or \textit{forecast ensemble} starting from this single solution.
As before, we can define the forward flow map as the (stochastic) map $\phi_{\Delta t}(w^n):= w^{n+1}=w^n + \int_{t^n}^{t^{n+1}}\! L_\mathrm{LES}(w^n)+m(w^n)\,\td t$, where $m(w^n)$ is a stochastic LES model.
Thus, through the choice of numerical method and LES model, we implicitly define the transition kernel $\tau_{\mathrm{LES},\Delta t}(w_f|w)$. 
This kernel can be thought of as describing the probability that $\phi_{\Delta t}(w)=w_f$ occurs, or in other words, that the flow configuration $w_f$ is reached after integrating the prediction model $\Delta t$ time units starting from configuration $w$. 
Using this notation, we define the forecast distribution $\pi_f^{n+1}$ at time $t^{n+1}$ as \begin{equation}
    \pi_f^{n+1}(w_f) = \int_\MW\! \tau_{\mathrm{LES},\Delta t}(w_f|w)\delta(w-w^n)\,\td w.
\end{equation}

If one defines the stochastic LES model $m(w)$ such that its average equals the conditional average in equations (\ref{eq:ideal_model}-\ref{eq:ideal_measurement}), then the mean of the stochastic ensemble will be the ideal LES prediction.
However, in practice this will generally not be achievable.
The conditional average in equation \eqref{eq:ideal_model} can be estimated with sufficient measurements of $M(u)$ (equation \eqref{eq:ideal_measurement}) using nonlinear mean square estimation \cite{papoulis2002probability}. 
Nonetheless, it becomes increasingly difficult to obtain sufficient samples of the state space when its dimension increases \cite{reich2015probabilistic}, hence it is infeasible to compute good approximations of the conditional average even at modest resolutions of the LES prediction. 
In practice, therefore, the forecast distribution $\pi_f^{n+1}$ will generally differ from the ideal distribution and may even be markedly different if errors accumulate. 
We thus need to correct the forecast distribution based on available knowledge of the ideal distribution.

\paragraph{Bayesian correction to incorporate measurements}
There is no guarantee a produced forecast ensemble is an accurate approximation of the ideal distribution.
We incorporate known information of the ideal distribution $\pi_W^{n+1}$ into the forecast distribution $\pi_f^{n+1}$ in a Bayesian assimilation step, thereby producing an \textit{analysis ensemble} or \textit{posterior ensemble} with a distribution denoted by $\pi_a^{n+1}$.
The assimilation can succinctly be written in terms of a transition kernel $b(w_a|w_f)$ acting on the forecast ensemble \cite[Chapter 7]{reich2015probabilistic}, \begin{equation}
    \pi_a^{n+1}(w_a) = \int_\MW\! b(w_a|w_f)\pi_f^{n+1}(w_f)\,\td w_f,
\end{equation}
where the form of $b$ depends on the adopted DA method.

As an example we consider the Ensemble Kalman filter (EnKF) with perturbed observations \cite{evensen1994sequential, evensen2003ensemble} as a corrector for the prediction ensemble.
EnKF provides a practical algorithm since it defines the transition from a forecast distribution to an analysis distribution in terms of the corresponding ensembles.
We denote by $w_{i, f}^{n+1}$ and $w_{i, a}^{n+1}$ the $i^\mathrm{th}$ ensemble members of the forecast ensemble and analysis ensemble, respectively, at time $t^{n+1}$.
The correction then takes the form \begin{equation}
    w_{i, a}^{n+1} = w_{i, f}^{n+1} - K\left(H w_{i, f}^{n+1} + r_i^{n+1} - y_\mathrm{obs}^{n+1} \right), 
\end{equation}
where $r_i^{n+1}\sim\MN(0, R)$ and $y_\mathrm{obs}^{n+1}$ is an observation. 
Here, $H$ is the measurement operator mapping an LES prediction $w_{i, f}^{n+1}$ to an observable, $K$ is the Kalman gain matrix and $R$ is the observational noise covariance matrix. For this method, the transition kernel $b$ can be expressed as \cite[Section 7.1]{reich2015probabilistic} \begin{equation}
    b(w_a^{n+1}|w_f^{n+1}) = \mathrm{n}(w_a^{n+1}; w_f^{n+1} - K(Hw_f^{n+1} - y_\mathrm{obs}^{n+1}), KRK^T).
\end{equation}
The term on the right-hand side describes the probability of sampling $w_a^{n+1}$ from a normal distribution with mean $w_f^{n+1} - K(Hw_f^{n+1} - y_\mathrm{obs}^{n+1})$ and covariance $KRK^T$.

\subsection{Closure modelling by assimilating statistics}
\label{subsec:general_closure}
Using the assumption that the flow is statistically stationary and that high-fidelity flow snapshot data is available, we now derive a framework for ensemble methods that serve as a self-contained data-driven stochastic models.

A flow statistic is defined by a function $G:\MW\to\R$ producing a single number for a given flow configuration.
Then, the expected value of the statistic in the ideal distribution \begin{equation}
    \BE_{\pi_W}\left[G(w)\right] = \int_\MW\!G(w)\pi_W(w)\,\td w
\end{equation}
becomes time-independent since we assume statistical stationarity. 
Therefore, instead of focusing on approximating the ideal distribution $\pi_V^{n+1}$ by the posterior distribution $\pi_a^{n+1}$, we can choose to relax this modelling goal such that it applies to selected flow statistics only.
The underlying assumption is that selected flow statistics adequately describe the ideal distribution and hence that an ensemble of solutions with accurate flow statistics approximates the ideal distribution.
That is, we aim to find a posterior distribution $\pi_a^{n+1}$ of flow fields such that \begin{equation}
    \BE_{\pi_a^{n+1}}\left[G(w)\right] =\BE_{\pi_W}\left[G(w)\right]
    \label{eq:statistics_agreement}
\end{equation}
for user-defined flow statistics $G$.
This way, we shift our focus from predicting distributions of fields to predicting distributions of flow statistics.
In doing so, we exploit the knowledge of the distribution of $G$ as measured from high-fidelity flow snapshots.
We denote this distribution by $G(w)\pi_W$.

To achieve good approximations of flow statistics, following \eqref{eq:statistics_agreement}, we consider sequential data assimilation applied to a predicted ensemble of statistics rather than an ensemble of flow fields.
This gives rise to the following general method, also summarized in algorithm \ref{alg:framework_general}.
Starting from an initial field $w^n$, an LES prediction ensemble $\{w_{i,f}^{n+1}\}$ is computed from which flow statistics $\{G_{i, f}^{n+1}\}$ are extracted. 
A Bayesian method is subsequently employed to correct the predicted statistics.
Namely, a DA method is applied to the predicted statistic ensemble $\{G_{i, f}^{n+1}\}$, where `observations' $\{G_{i, \mathrm{obs}}^{n+1}\}$ are samples from the distribution $G(w)\pi_W$. 
The latter is estimated from the high-fidelity data.
The resulting updated values $\{G_{i, a}^{n+1}\}$ of the statistic should be satisfied by the analysis ensemble of flow fields $\{w_{i, a}^{n+1}\}$.
Thus, the final step of the algorithm is to find, or `reconstruct', this ensemble.
Specifically, each forecast ensemble member $w_{i, f}^{n+1}$ is transformed into an analysis ensemble member $w_{i, a}^{n+1}$ such that the latter satisfies the flow statistic $G_{i,a}^{n+1}$.

\begin{algorithm}
    \caption{Framework for closure modelling by assimilating statistics over one time step}
    \label{alg:framework_general}
    \begin{algorithmic}
        \Procedure{Closure Model}{$w^n$, $L_\mathrm{LES}$, $\Delta t$, $N$\textbf{more ... }}

        \For{$i=1, \ldots, N$} 
        \State $w_{i, f}^{n+1} \gets w^n + \int_{t^n}^{t^{n+1}}\! L_\mathrm{LES}(w^n)+m(w^n)\,\td t$
        \Comment{Compute ensemble LES prediction}
        \State $G_{i, f}^{n+1} \gets G(w_{i, f}^{n+1})$
        \Comment{Compute predicted statistics}
        \State $G_{i, \mathrm{obs}}^{n+1} \gets \mathrm{sample}(G(w)\pi_W)$
        \Comment{Generate `observed' statistics}
        \EndFor

        \State $\left\{G_{i, a}^{n+1}\right\}\gets \mathrm{DA}\left(\left\{G_{i, f}^{n+1}\right\}, \left\{G_{i, \mathrm{obs}}^{n+1} \right\}\right)$
        \Comment{Bayesian correction of predicted statistics}

        \For{$i=1, \ldots, N$} 
        \State $w_{i, a}^{n+1} \gets \mathrm{reconstruct}(w_{i, f}^{n+1}, G_{i, a}^{n+1})$
        \Comment{Reconstruct flow fields such that these satisfy the updated statistics}
        \EndFor
        \State \Return $\left\{w_{i, a}^{n+1}\right\}$
        \Comment{Return the ensemble of flow fields with updated statistics}
        \EndProcedure
    \end{algorithmic}
\end{algorithm}

Algorithm \ref{alg:framework_general} provides a general framework for including knowledge of desired flow statistics into stochastic LES predictions. 
However, its generality comes at the cost of having many modelling choices, which we list below.
These choices are
\begin{enumerate}
    \item how to measure the sub-grid force $M(u)$ in equation \eqref{eq:ideal_measurement}. 
    Ideal LES establishes what this measurement should encompass in continuous time. 
    In discrete time, one has to decide over which time interval $M(u)$ is measured;
    \item which stochastic LES model to use.
    Even though ideal LES provides a goal for what the LES model should achieve, the form of the stochastic forcing term may depend on the problem at hand;
    \item which statistics to assimilate; 
    \item how the `observations' are defined based on high-fidelity data. 
    Time series data of statistics can be extracted from available high-fidelity flow snapshots.
    In turn, one can choose, e.g., to draw samples from empirically measured distributions, or to mimic measured temporal correlation;
    \item which DA scheme to use;
    \item how to reconstruct the flow fields when statistical values are known. A small number of statistics does not uniquely define a flow field, hence it is desirable to optimally find flow fields $w_{i,a}^{n+1}$ from $w_{i,f}^{n+1}$ under the constraints defined by the desired flow statistics.
\end{enumerate}

The first two points presented above are related to ensemble prediction with stochastic data-driven LES, the remainder are part of the Bayesian correction.
In the next section, we illustrate the presented closure modelling framework for two-dimensional \RB convection and highlight each of the required modelling choices.

\section{Application to Rayleigh-B\'enard convection}\label{sec:application}
In this section, we introduce the governing equations and the test case that the closure model will be applied to. 
Subsequently, we elaborate on the model choices, provide implementation details and an estimate of the resulting computational complexity. 

\subsection{Governing equations and numerical methods}\label{subsec:rb_eqns}
Two-dimensional \RB (RB) convection serves as the test bed for the proposed model. The governing equations are the incompressible Navier-Stokes equations coupled to an energy equation describing buoyancy effects under the Boussinesq approximation. The nondimensionalized equations are
\begin{align}
    \frac{\partial \mathbf{u}}{\partial t} + \mathbf{u} \cdot \nabla\mathbf{u} &= \frac{\mathit{Pr}}{\mathit{Ra}}\nabla^2 \mathbf{u} - \nabla p + T\mathbf{e}_y, \label{eq:rb_momentum}\\
    \nabla\cdot \mathbf{u} &= 0, \label{eq:rb_cont} \\
    \frac{\partial T}{\partial t} + \mathbf{u}\cdot\nabla T &= \frac{1}{\sqrt{\mathit{Pr} \mathit{Ra}}}\nabla^2 T, \label{eq:rb_energy}
\end{align}
where $\mathbf{u}$ is the velocity, $p$ is the pressure, $T$ is the temperature, and $\mathbf{e}_y$ is the unit vector in the vertical direction. 
The velocity consists of a horizontal velocity $u_x$ and a vertical velocity $u_y$.
Following the notation of Section \ref{sec:ideal_les}, we regard a solution $u$ as the pair $(\mathbf{u}, T)$.

The equations are solved in a two-dimensional rectangular box of width $L_x=2$ and height $L_y=1$. 
Periodic boundary conditions are imposed on the sides of the domain for all variables. 
The top and bottom boundaries are walls with no-slip boundary conditions for the velocity, and prescribed values of 1 at the bottom and 0 at the top for the dimensionless temperature. 
The dimensionless numbers are the Prandtl number $\mathit{Pr}=\nu / \kappa$ and the Rayleigh number $\mathit{Ra}=g\beta\Delta L_y^3/(\nu\kappa)$. 
These numbers describe the ratio of characteristic length scales of the velocity and the temperature and the ratio between buoyancy and viscous effects, respectively, and we have a Reynolds number $\mathit{Re}=\sqrt{\mathit{Ra}/\mathit{Pr}}$. 
Here $\nu$ is the kinematic viscosity, $\kappa$ is the thermal diffusivity, $g$ is the gravitational acceleration, $\beta$ is the thermal expansion coefficient, $\Delta$ is the temperature difference between the walls of the domain.
The test case is run at $\mathit{Pr}=1$ and $\mathit{Ra}=10^{10}$ to simulate in the turbulent convective regime.

The Nusselt number describes the heat flux from the bottom to the top of the domain and is one of the critical responses of the system to the imposed physical parameters \cite{ahlers2009heat}. 
We adopt the definition \begin{equation}
    \mathit{Nu} = 1 + \sqrt{\mathit{Pr}\mathit{Ra}} \langle vT \rangle_\Omega, \label{eq:definition_nusselt}
\end{equation}
where $\langle\cdot\rangle_\Omega$ is the mean over the entire domain $\Omega$. 
Definition \eqref{eq:definition_nusselt} is particularly suited for computation on coarse grids since it does not involve any gradients, and will be used as a metric to assess coarse-grid simulations.

The equations (\ref{eq:rb_momentum}-\ref{eq:rb_energy}) are discretized in space using a energy-conserving finite difference method \cite{vreman2014projection}.
The high-fidelity solver is parallelized following \cite{cifani2018highly}. 
The velocity is defined on a staggered grid, the temperature is defined on the same grid as the vertical velocity, and the pressure is computed at the cell centers. 
The arrangement of the temperature and vertical velocity ensures no interpolation errors when computing the buoyancy term \cite{van2015pencil}. 
A hyperbolic tangent grid spacing is adopted in the vertical (wall-normal) direction, guaranteeing refinement near the walls to resolve the boundary layer.
The grid is uniformly spaced in the horizontal direction.

The temporal discretization follows from a fractional-step third-order Runge-Kutta (RK3) for explicit terms and the Crank-Nicholson (CN) scheme for implicit terms. 
A time step from $t^n$ to $t^{n+1}$ is divided into three sub-stages denoted by the superscript $k, k=0, 1, 2$, where $k=0$ coincides with the situation at $t^n$. 
In each stage, a provisional velocity $\mathbf{u}^*$ is computed as \begin{equation}
    \frac{\mathbf{u}^* - \mathbf{u}^k}{\Delta t} = \left[ \gamma_k H^k + \rho_k H^{k-1} - \alpha_k \mathcal{G}p^k + \alpha_k\mathcal{A}_y^k \frac{\mathbf{u}^* +\mathbf{u}^k}{2} \right].
\end{equation}
The Runge-Kutta coefficients $\gamma, \rho, \alpha$ are given by $\gamma=[8/15, 5/12, 3/4], \rho=[0, -17/60, -5/12]$ and $\alpha=\gamma+\rho$ (see \cite{rai1991direct, van2015pencil, cifani2018highly}). 
The discrete gradient is denoted by $\mathcal{G}$. 
The convective terms, horizontal diffusion terms and source terms (buoyancy) are collected in $H^k$ and are treated explicitly. 
The vertical diffusion term $\mathcal{A}_y$ is treated implicitly to eliminate viscous stability restrictions arising from the non-uniform grid near the boundary \cite{kim1985application}. 
A Poisson equation is then solved using $\mathbf{u}^*$ to impose the continuity constraint \eqref{eq:rb_cont}. 
Discretely, this is given by \begin{equation}
    \mathcal{L}\phi = \frac{1}{\alpha_k \Delta t}\left(\mathcal{D}\mathbf{u}^* \right), \label{eq:poisson}
\end{equation}
where $\mathcal{L}$ is the discrete Laplacian and $\mathcal{D}$ is the discrete divergence. 
The velocity and pressure are then updated to yield a divergence-free velocity field, \begin{align}
    \mathbf{u}^{k+1} & = \mathbf{u}^* - \alpha_k \Delta t(\mathcal{G}\phi), \\
    p^{k+1} &= p^k + \phi - \frac{\alpha_k\Delta t}{2\mathit{Re}}\left(\mathcal{L}\phi\right).
\end{align}
The QUICK interpolation scheme \cite{leonard1979stable} is used to discretize the convective terms, whereas the diffusive terms are discretized with a standard second-order finite difference scheme in both spatial directions. 
Finite differences are used to define the discrete gradient $\mathcal{G}$, the discrete divergence $\mathcal{D}$ and the discrete Laplacian $\mathcal{L}$.

A resolution of $4096\times 2048$ is adopted for the DNS, which has been shown to be sufficient to resolve the flow at the chosen Rayleigh number \cite{zhu2018transition}. 
All subsequent coarse-grid simulations are carried out on a $64\times 32$ grid. 
The flow cannot be fully resolved at this low resolution and the numerical method induces artificial dissipation \cite{geurts2005numerically, ephrati2023rb}. 
Hence, the low-fidelity numerical solution will reach a different statistically steady state than the high-fidelity reference solution if no model is explicitly applied.

\subsection{Model specification} \label{subsec:rb_model}
As presented in algorithm \ref{alg:framework_general} in Section \ref{sec:model}, 
the statistical closure modelling framework consists of a stochastic ensemble forecast method and a subsequent Bayesian correction.
A total of six modelling choices have to be made to obtain a workable algorithm, which we elaborate below.

We assume that a sequence of high-fidelity snapshots $\{u^j\}$ is available.
Furthermore, we repeatedly make use of the periodicity of the domain in the horizontal (wall-parallel) direction by computing the one-dimensional Fourier transform along horizontal cross-sections of the domain.
The Fourier coefficients are indicated with a hat symbol $\hat{\cdot}$.
For example, we use $\widehat{u_x}_{,k, l}$ to denote the $k^\mathrm{th}$ Fourier coefficient of the velocity $u_x$ along the $l^\mathrm{th}$ horizontal cross-section.
Given an entire solution $u$ consisting of a velocity $\mathbf{u}=(u_x, u_y)$ and a temperature $T$, one obtains the Fourier coefficients $\widehat{u_x}_{, k, l}$, $\widehat{u_y}_{, k, l}$, $\widehat{T}_{k, l}$.
We denote these coefficients collectively by $\widehat{u}_{k, l}$ for readability.
This leads to the following model description.

\begin{enumerate}
    \item The subgrid force $M$ in equation \eqref{eq:ideal_measurement} is recorded in an \textit{a posteriori} measurement.
    Specifically, $M(u^n)$ is measured by integrating the high-fidelity solver in time from a high-fidelity snapshot $u^n$ and the low-fidelity solver from $\tilde{u}^n$.
    Both solutions are integrated for a single time step, after which the solution from the high-fidelity solver is filtered and compared to the solution of the low-fidelity solver.
    Thus, a set of measurements $\{M(u^j)\}$ is obtained from a collection of high-fidelity snapshots $\{u^j\}$.
    This procedure is summarized in algorithm \ref{alg:a_post_measurement}.
    The number of measurements is varied throughout the performed numerical experiments and is specified in Section \ref{sec:results}.
\begin{algorithm}
    \caption{Single a posteriori measurement}
    \label{alg:a_post_measurement}
    \begin{algorithmic}
        \Procedure{A posteriori measurement}{$u^n$, $L$, $L_\mathrm{LES}$, $\tilde{\cdot}$, $\Delta t$}
        \State $w^n \gets \tilde{u}^n$ 
        \Comment{Initialize low-fidelity solution as the filtered high-fidelity snapshot}
        \State $u^{n+1} \gets \int_{t^n}^{t^{n+1}} L(u^n)\,\td t $
        \Comment{Integrate high-fidelity solution in time}
        \State $w^{n+1} \gets \int_{t^n}^{t^{n+1}}L_\mathrm{LES}(w^n)\,\td t$
        \Comment{Integrate low-fidelity solution in time}
        \State $M(u^n) = \tilde{u}^{n+1} - w^{n+1}$
        \State \Return $M(u^n)$
        \EndProcedure
    \end{algorithmic}
\end{algorithm}

\item The LES model is implemented as a stochastic perturbation calibrated from the measurements $\{M(u^j)\}$.
Namely, time series of the Fourier coefficients along horizontal cross-sections of the domain are first extracted for each prognostic variable. 
This yields $\{\widehat{M(u^j)}_{k, l}\}$ and in particular we also obtain magnitudes $\{\mathrm{abs}(\widehat{M(u^j)}_{k, l})\}$.
The sample means $\mu_{M, k, l}$ and variances $\sigma^2_{M, k, l}$ of the magnitudes are stored.

The perturbation is then constructed as follows.
For each Fourier coefficient and horizontal cross-section independently, we sample a magnitude $r_{M, k, l}\sim\MN(\mu_{M, k, l}, \sigma^2_{M, k, l})$ and a phase $\phi_{M, k, l}\sim U[0, 2\pi]$.
This fully defines the spectral coefficients and thus the corresponding physical fields. 
These fields are added as a perturbation after the full time step has been completed with the low-fidelity solver.
An ensemble forecast $\{w_{i,f}^{n+1}\}$ is produced by computing a separate perturbation for each ensemble member. 
Such a perturbation can accurately compensate for coarsening errors even when a fraction of data is used \cite{ephrati2022computational} and is non-intrusive, thereby readily enabling ensemble forecasts. 

\item We aim to reproduce the energy spectra and mean heat flux in all horizontal cross-sections of the domain. 
Given a solution $u$, the QoIs are $\mathrm{abs}(\widehat{u}_{k, l})$ for the energy spectra and $(\widehat{u_yT})_{0, l}$ for the mean heat flux, where the subscript $0$ signifies the zeroth Fourier coefficient.
For simplicity, we will denote a single QoI by $G$, which can refer to either the magnitude of a Fourier coefficient or the mean heat flux.

\item The `observations' are based on the sequence of filtered high-fidelity snapshots $\{\tilde{u}^j\}$. 
This sequence yields a time series $\{G^j_\mathrm{ref}\}$ for each QoI, of which the sample mean $\mu_\mathrm{obs}$ and variance $\sigma_\mathrm{obs}^2$ are stored.
At every assimilation step, an ensemble $\{G_{i,\mathrm{obs}}\}$ of observations is generated by sampling $G_{i, \mathrm{obs}}\sim\MN(\mu_\mathrm{obs}, \sigma_\mathrm{obs}^2)$ for each QoI separately.

\item A simplified ensemble Kalman filter (EnKF) \cite{evensen1994sequential} is employed to update the predicted statistics based on the observations. 
EnKF takes into account the nonlinearity of the governing equations, permits using a wide range of noise models \cite{evensen2003ensemble}, and is straightforward to implement. 

A diagonalization approach \cite{harlim2008filtering, majda2012filtering} is used that disregards covariances between the QoIs, so that the analysis acts on each QoI separately. 
This can be thought of as covariance `localization'.
Consequently, this avoids computing (inverses of) covariance matrices in the analysis step as is usually required in EnKF, which becomes a source of extensive computational costs if many statistics are assimilated simultaneously. 
Instead, the analysis for each QoI reduces to a scalar equation independent of other QoIs. 
No ensemble inflation is used in the present study.

We denote the forecast, analysis ensemble, and observation ensembles of a single QoI by $\{G_{i, f}\}$, $\{G_{i, a}\}$, and $\{G_{i, \mathrm{obs}}\}$, respectively.
The EnKF then updates each predicted QoI as \begin{equation}
    G_{i, a} = G_{i, f} + \left(\frac{\mathrm{var}(\{G_{i, f}\})}{\mathrm{var}(\{G_{i, f}\}) + \mathrm{var}(\{G_{i, \mathrm{obs}}\})}\right) \left(G_{i, \mathrm{obs}}-G_{i,f}\right).
    \label{eq:EnKF_diagonal}
\end{equation}


\item The predicted LES solution $w_{i, f}$ is updated to a solution $w_{i,a}$ satisfying the statistic $G_{i,a}$ from the previous step by appropriately adjusting the spectral coefficients of $w_{i, f}$.
Namely, the magnitudes of the spectral coefficients of $w_{i, f}$ can be readily changed to $G_{i, a}$. 
The desired heat flux values are approximated by applying a gradient descent method on the phases of the Fourier coefficients of the temperature field, as developed in \cite{ephrati2023rb}.
The solution $w_{i, a}$ is then fully described by the updated spectral coefficients.
\end{enumerate}

The diagonalized EnKF in step 5 updates each QoI separately by disregarding covariances between the QoIs.
Consequently, the error $|G_{i, a}-G_{i, \mathrm{obs}}|$ can be estimated for each QoI independently, and is further detailed in Appendix \ref{app:error_estimates}.
In particular, for a sufficiently large ensemble size and sufficiently small time step, the ensemble mean of $\{G_{i,a}\}$ is shown to revert to $\muo$.
Accurately predicting mean values of QoIs is a consistency requirement, since it is assumed that the specified QoIs adequately describe the ideal distribution.

\subsection{Computational complexity}
We now provide some estimates of the complexity reduction when employing the model on coarse grids. 
Solving the Poisson equation \eqref{eq:poisson} is the largest source of computational costs in the spatial discretization since it involves solving a sparse $(N_x N_y)\times(N_x N_y)$ linear system. 
The high-fidelity solver employs a parallelization technique for finite-difference discretization of wall-bounded flows \cite{van2015pencil} which solves \eqref{eq:poisson} in $\MO(N_x N_y \log[N_y])$ operations. 
The low-fidelity solver uses a standard LU factorization for sparse linear systems for this purpose, requiring $\MO((N_x N_y)^{3/2})$ computational steps \cite{george1988complexity}. 

The one-dimensional (inverse) fast Fourier transform (I(FFT)) along horizontal cross-sections of the domain is used extensively in the present adaptation of the model. 
These are computed in $\MO(N_x \log[N_x])$ operations \cite{golub2013matrix}. 
The random fields in the stochastic perturbations are computed by applying the IFFT at each horizontal profile for all prognostic variables. 
Applying the correction of the magnitudes of the Fourier coefficients requires computing the (I)FFT twice for these variables. 
The heat flux correction requires three computations of the (I)FFT in total and an additional $\MO(N_x N_y)$ arithmetic operations per iteration. 
In total, we obtain the estimates of $\MO(10^7)$ operations per time step for the high-fidelity solver, and  $\MO(10^5)$ for the low-fidelity solver including the model.
This estimate is for a single realization of the model.
Simulating an ensemble of size $N$ increases the computational costs by approximately a factor $N$.
The diagonalized EnKF \eqref{eq:EnKF_diagonal} does not result in substantial additional computational costs.

The time step size becomes restrictive for high-$\mathit{Ra}$ flow. 
A step size of $\MO(\Delta y)\approx \MO((\mathit{Ra}\mathit{Nu})^{-1})=\MO(10^{-12})$ is suggested \cite{van2015pencil} to stably fully resolve the turbulent flow features. 
Here $\Delta y$ denotes the smallest grid spacing in the non-uniform wall-normal grid. 
The employed fine and coarse grids differ by a factor 64 in terms of grid cells per spatial direction.
However, the difference between the smallest (non-uniform) grid spacings in the wall-normal direction yields a significant relaxation of the time step constraint.
Following the estimate above, a time step restriction of $\MO(10^{-4})$ is obtained on the adopted coarse grid.
However, a time step size of $\MO(10^{-2})$ was found to yield stable results, likely due to the presence of artificial dissipation, leading to a significant computational cost reduction. 

\section{Model performance assessment}\label{sec:results}
This section presents simulation results on the coarse computational grid, comparing four models to the high-fidelity reference. 
We first introduce the employed coarse-grid models, the reference data used for model calibration, and the performance assessment metrics.

\paragraph{Coarse-grid models}
Four different models are employed in the coarse-grid simulations and compared to the filtered DNS.
These models are listed below and consist of a physics-based deterministic model and three ensemble methods.
\begin{itemize}
    \item The no-model coarse numerical simulation serves as a deterministic physics-inspired closure model and will be referred to as the `\textit{coarse, no model}' method.
    The artificial viscosity native to the coarse discretization was found to produce results similar to using a Smagorinsky eddy-viscosity model \cite{smagorinsky1963general, peng1998comparison} at several Smagorinsky constants.
    The latter is not included in the results for brevity.
    
    \item The first ensemble method is a stochastic LES model without Bayesian correction, which we refer to as a `\textit{random sgs}' (subgrid-scale) model.
    The model uses only the first two steps described in Section \ref{subsec:rb_model}. Hence, it relies only on the a posteriori measurements and does not require additional time series data of the QoIs.
    This model is a variant of the a posteriori statistical turbulence closure of \cite{hoekstra2024reduced}, applied to the quantities of interest given in Section \ref{subsec:rb_model}.
    
    \item The second ensemble method combines the coarse no-model simulation with a heuristic stochastic correction, and is referred to as a `\textit{statistical nudge}'.
    Namely, an ad-hoc correction of the predicted QoIs is applied towards an observation following the work of \cite{ephrati2023rb}.
    The observation is defined as in the third step in Section \ref{subsec:rb_model}, while the nudging strength is defined for each statistic independently through the measured correlation time \cite{ephrati2023rb}.
    This approach does not use the a posteriori measurements but instead only requires time series data of the QoIs.
    
    \item The third ensemble method combines stochastic LES with a Bayesian correction, using all steps described in Section \ref{subsec:rb_model}, and will be denoted as the `\textit{random sgs, assimilated}' model.
\end{itemize}
The ensembles will consist of 10 members in Sections \ref{subsec:plentydata} and \ref{subsec:fewdata}.
An ensemble with 50 members is considered in Section \ref{subsec:ensemblesize} to investigate the dependence of the prediction quality on the ensemble size.

\paragraph{Reference data}
The reference data is a sequence of filtered snapshots of a DNS performed on a $4096\times 2048$ computational grid, from which two data sets are extracted.
\begin{itemize}
    \item The first data set is comprised of `\textit{plenty data}'.
    A total of 1000 solution snapshots separated by 0.05 time units are used to compute the a posteriori measurements and compute the statistics used in the observations.
    The corresponding results are presented in Sections \ref{subsec:plentydata} and \ref{subsec:ensemblesize}.
    \item The second data set consists of `\textit{few data}' and is used to verify the robustness of the model in the sparse data regime.
    Here, a total of 20 solution snapshots separated by 0.5 time units are used in the model calibration.
    The results using this data set are presented in Section \ref{subsec:fewdata}.
\end{itemize}
The initial conditions of all coarse-grid simulations are filtered DNS snapshots outside the data sets described above.


\paragraph{Performance metrics}
Several metrics are used to assess the model performance. 
For short lead times, a predicted solution $(\mathbf{u}, T)$ can be compared to the reference solution $(\mathbf{u}_\mathrm{ref}, T_\mathrm{ref})$ using the pattern correlation.
We adopt the definition \begin{equation} 
\frac{\left\langle (\mathbf{u}, T), (\mathbf{u}_\mathrm{ref}, T_\mathrm{ref})  \right\rangle}{\sqrt{\left\langle (\mathbf{u}, T), (\mathbf{u}, T) \right\rangle \left\langle (\mathbf{u}_\mathrm{ref}, T_\mathrm{ref}), (\mathbf{u}_\mathrm{ref}, T_\mathrm{ref}) \right\rangle} }, \label{eq:pattern_correlation}
\end{equation}
where $\langle \cdot, \cdot\rangle$ denotes the standard $L^2$ inner product over the computational domain. 

Long-time simulation results are evaluated via time-averaged energy spectra and root-mean-square deviations (r.m.s.), as well as rolling means of the kinetic energy (KE) and the Nusselt number. 
The r.m.s. is computed as a function of the wall-normal distance. For a given $y$-value, we compute the r.m.s. of a field $f$ as \begin{equation}
    \mathrm{r.m.s. }(f, y, t) = \left[\frac{1}{|A|}\int_A\! \left(f(x, y, t) - \langle f(x, y, t) \rangle_A \right)^2\,\text{d}A \right]^{1/2}, 
\end{equation}
where $\langle\cdot\rangle_A$ is the mean over a cross-section with length $|A|$. The KE is defined as \begin{equation}
    \mathrm{KE} = \int_\Omega\! \frac{1}{2}\left(\mathbf{u}\cdot\mathbf{u}\right)\, \text{d}\Omega.
    \label{eq:definition_ke}
\end{equation}
The Nusselt number is computed following equation \eqref{eq:definition_nusselt}.

\subsection{Model performance with plenty data}\label{subsec:plentydata}
We first assess the model performance when plenty data is used to estimate the model parameters. 

\begin{figure}[h!]
    \centering
    \includegraphics[width=0.98\textwidth]{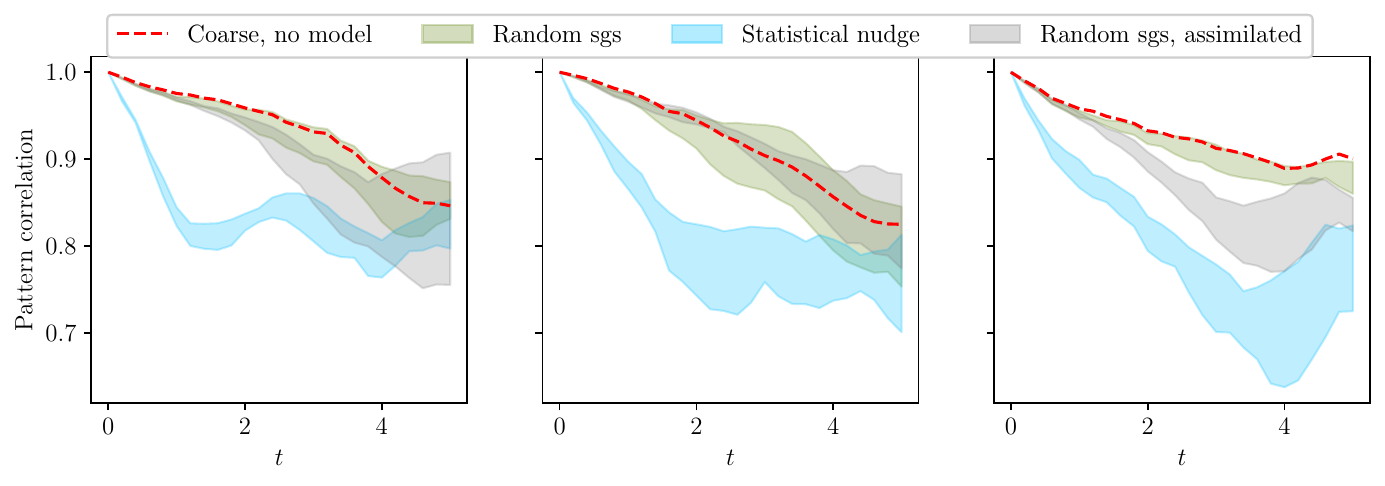}
    \caption{Pattern correlation between the prediction and the reference solution, using plenty data to calibrate the model. Three different initial conditions are considered. Each ensemble consists of 10 members; each band is colored between the maximal and minimal measured values.}
    \label{fig:plenty_pcorr}
\end{figure}

The pattern correlations \eqref{eq:pattern_correlation} at short lead times are shown in figure \ref{fig:plenty_pcorr} for three distinct initial conditions. The no-model simulation deteriorates over time due to loss of details in the numerical solution after initializing the flow from a filtered DNS snapshot. 
A low correlation does not imply that the flow statistics are incorrect since it measures the likeness of the global solution to the reference. Rather, it indicates the rate at which the flow deviates from the reference. It allows us to distinguish whether a solution is too rapidly steered towards a configuration with the desired statistical flow features.
The correlation of the statistical nudging ensemble with the reference decreases rapidly, from which we surmise that the ad-hoc nudging is too strong and does not lead to a physical flow evolution. Applying the stochastic perturbation largely mitigates this. The difference between the ensemble with random perturbations and the ensemble with perturbations and assimilated statistics suggests that a reduced nudging strength might be favorable when the initial condition is known. For example, this can be achieved by incorporating temporal effects into the observations, thereby including information of the initial conditions in the observation. Alternatively, a dynamically adaptive nudging procedure \cite{akyildiz2020nudging} can be applied such that a nudge is only applied if it increases the likelihood of the solution being in the ideal distribution.

\begin{figure}[h!]
    \centering
    \includegraphics[width=0.98\linewidth]{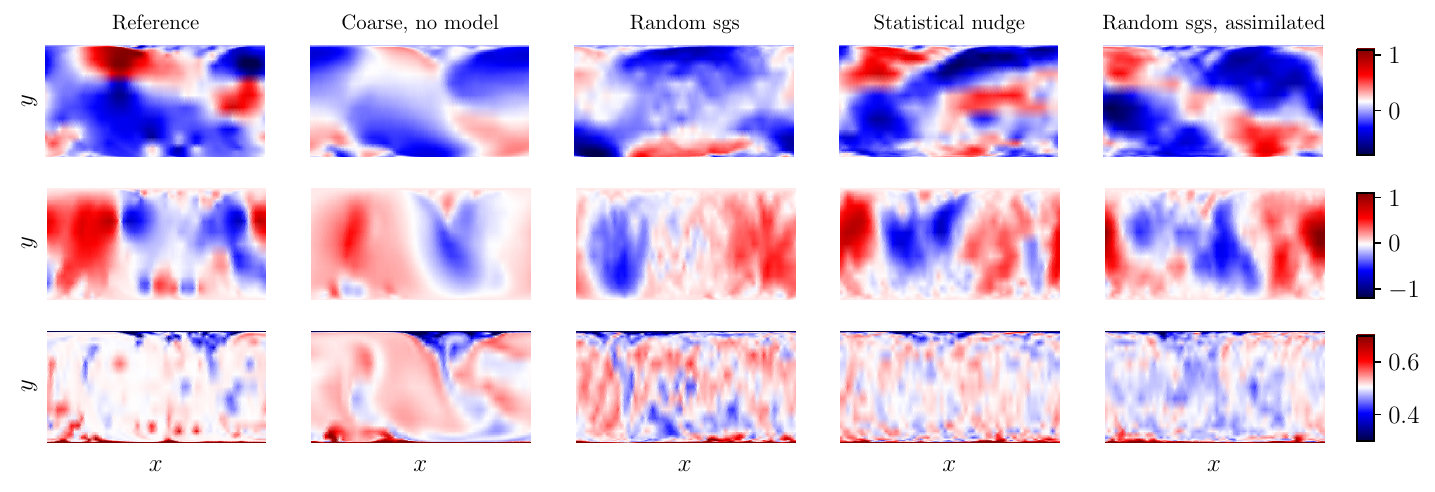}
    \caption{Instantaneous snapshots after simulating for 100 time units initialized from a filtered DNS snapshot. Shown are the horizontal velocity (top row), vertical velocity (middle row) and temperature (bottom row). A snapshot from a single ensemble member is shown for the ensemble methods.}
    \label{fig:plenty_snapshots}
\end{figure}

A qualitative model comparison is given in figure \ref{fig:plenty_snapshots} through instantaneous flow snapshots after reaching a statistically steady state in a long-time simulation. We observe that the no-model simulation suffers from artificial dissipation, evident from the smoothed fields and the reduced velocity magnitudes. The level of detail in the velocity fields is reconstructed well with the ensemble methods. The temperature fields display small-scale details, but these are fragmented instead of forming coherent patterns. Simultaneously, the velocity fields of the nudged ensembles feature pronounced flow details and maintain a qualitative agreement with the reference.

\begin{figure}[h!]
    \centering
    \includegraphics[width=0.98\linewidth]{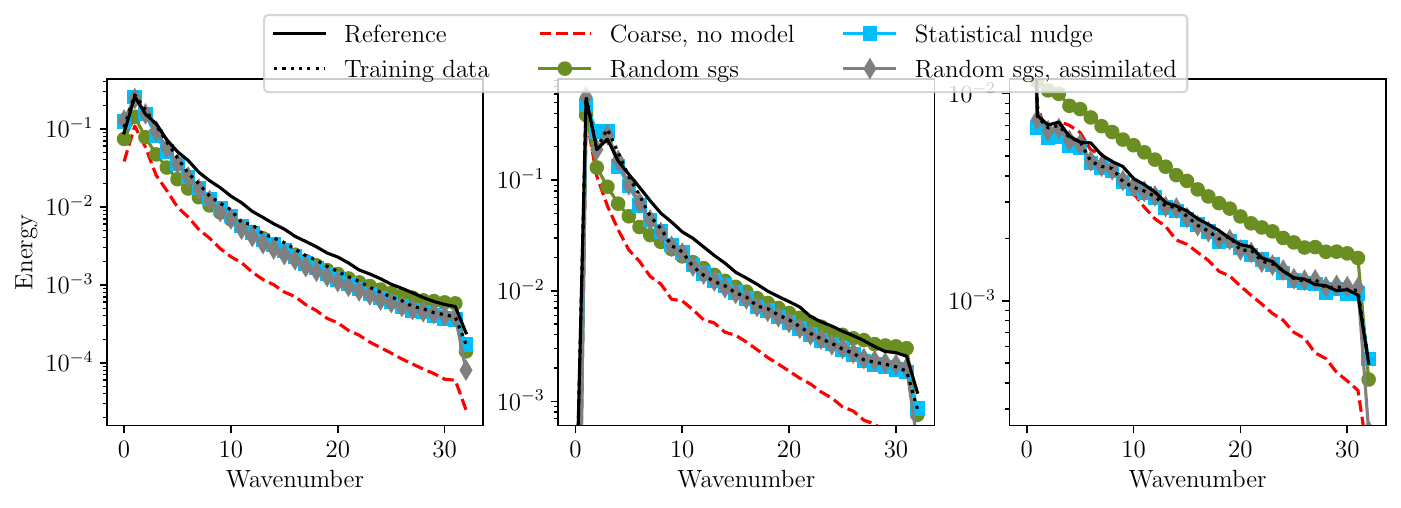}
    \caption{Time-averaged energy spectra measured along a horizontal cross-section of the domain for the horizontal velocity (left), vertical velocity (middle) and temperature (right). The cross-sections are take in the core of the domain at $y=5.5\times 10^{-1}$ for the horizontal velocity and $y=5.0\times 10^{-1}$ for the vertical velocity and the temperature.}
    \label{fig:plenty_spectra}
\end{figure}

The average energy spectra near the center of the domain are depicted in figure \ref{fig:plenty_spectra}, where the average is taken over all snapshots and all ensemble members. We also display the average spectra of the high-fidelity measurements that comprise the data set from which the model parameters are extracted and refer to this as the training data.
The no-model simulation provides predictions that consistently contain too little energy. Only applying the stochastic perturbation does not adequately alleviate this. Applying a correction, either ad-hoc or by assimilating statistics, yields a significant improvement particularly at the largest resolvable scales. This suggests that incorporating high-fidelity data into the model benefits the prediction of flow statistics in long-time simulations.
We observe that the resulting energy spectra approximate the training data accurately, which itself slightly deviates from the reference spectra over the simulated interval. 

\begin{figure}[h!]
    \centering
    \includegraphics[width=0.98\linewidth]{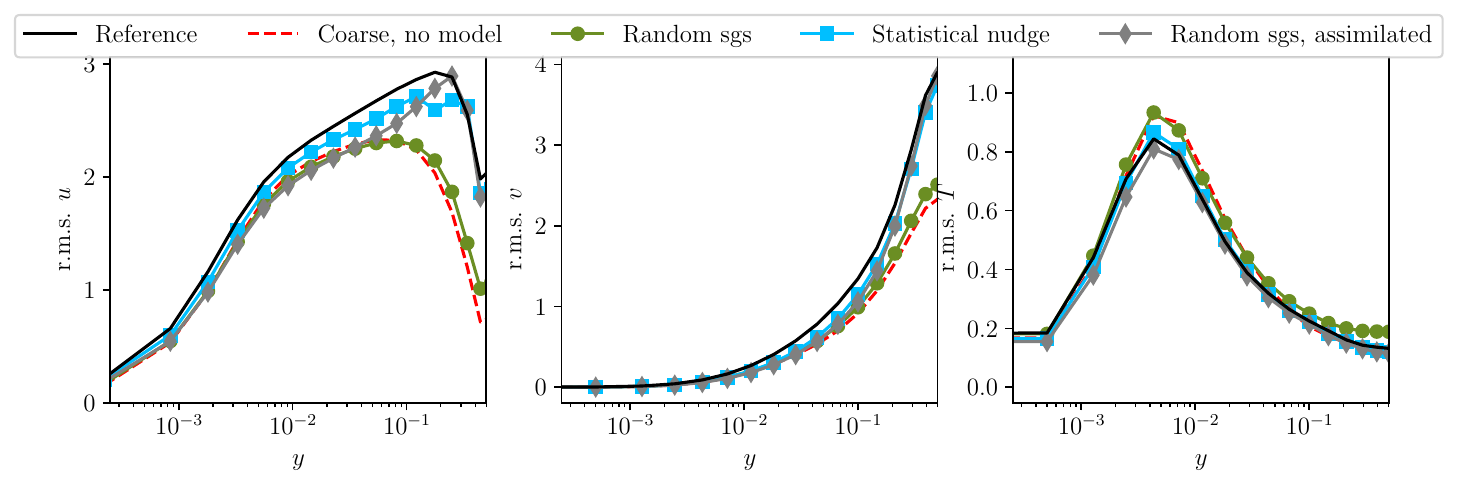}
    \caption{Average root mean square (r.m.s.) of the horizontal velocity (left), vertical velocity (middle) and temperature (right), measured along horizontal cross-sections of the domain and shown as functions of the wall-normal distance.}
    \label{fig:plenty_rms}
\end{figure}

The time-averaged r.m.s. values of show considerable improvement when applying the correction procedure, as shown in figure \ref{fig:plenty_rms}. Only applying the stochastic perturbation yields no improvement over using no model in the long-time simulations, which further highlights the added value of the statistical correction. Assimilating statistics leads to a pronounced improved of the velocity r.m.s. particularly near the center of the domain. Here, the grid size is largest and the discretization effects significant, which induce a large measured sub-grid force and thus large stochastic perturbations. In turn, these large perturbations lead to a larger gain factor in the correction \eqref{eq:enkf_diagonal} and a close adherence to the reference statistics.

\begin{figure}[h!]
    \centering
    \includegraphics[width=0.98\linewidth]{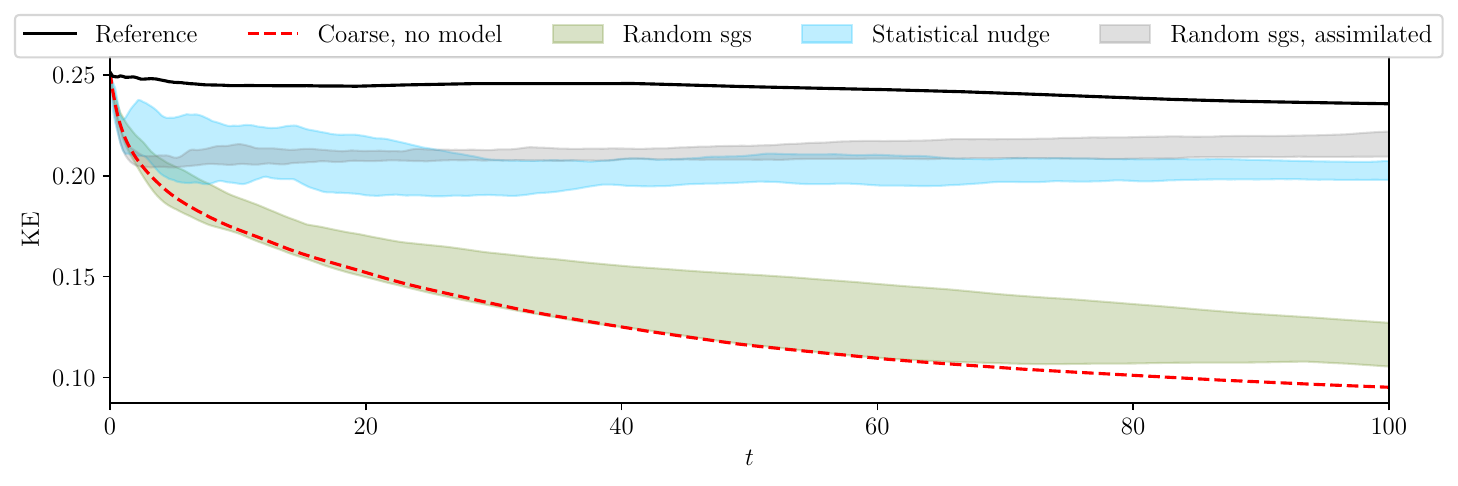}
    \caption{Rolling mean of the kinetic energy (KE) number over time.}
    \label{fig:plenty_kerm}
\end{figure}

\begin{figure}[h!]
    \centering
    \includegraphics[width=0.98\linewidth]{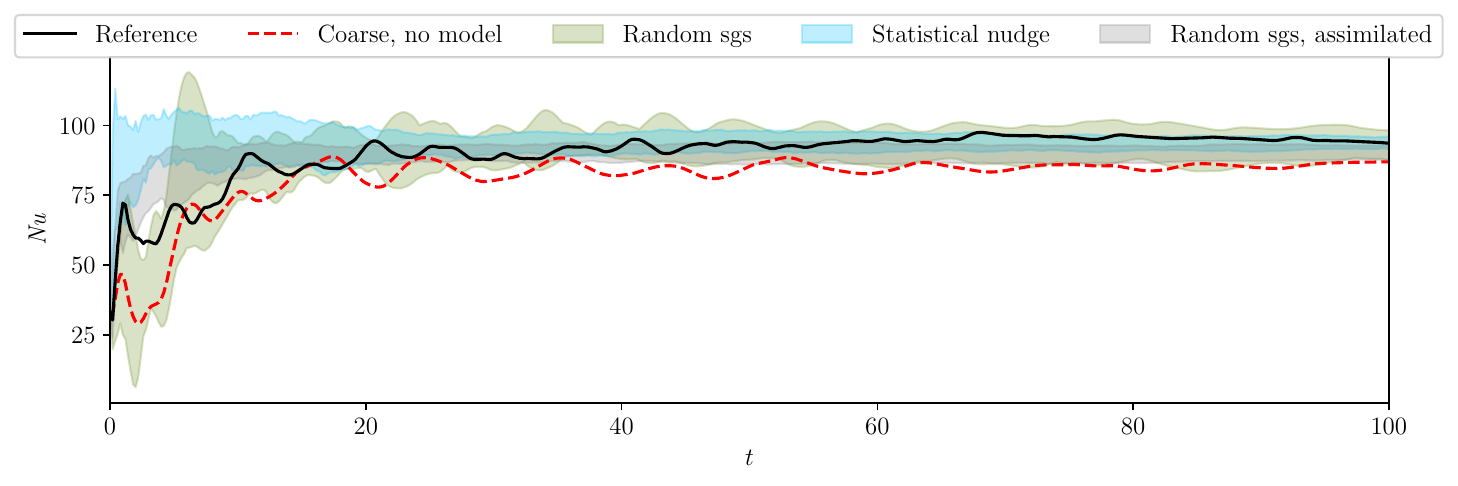}
    \caption{Rolling mean of the Nusselt number ($\mathit{Nu}$) over time.}
    \label{fig:plenty_nurm}
\end{figure}

The rolling averages of the total kinetic energy and the Nusselt number over time are respectively given in figures \ref{fig:plenty_kerm} and \ref{fig:plenty_nurm}. The artificial dissipation in the coarsened discretization causes the no-model result to deviate from the filtered DNS and reach a different statistically steady state with a strongly reduced energy content. Only applying the stochastic perturbation does not alleviate this, whereas including high-fidelity statistical data substantially improves the total energy content. An improvement of the predicted Nusselt number is observed for all ensemble methods. Notably, this includes the ensemble in which the predictions are perturbed and no knowledge of the heat flux is included. However, judging from the energy spectra in figure \ref{fig:plenty_spectra}, this result is obtained without correctly predicting the energy distributions in the velocity and temperature fields.


\subsection{Model performance with few data}\label{subsec:fewdata}
We now turn our attention to the model performance when using few data to estimate the model parameters. Only 20 snapshots are used to measure the the sub-grid scale forcing and the reference statistics. As such, the means and variances used in the model are poorly estimated and the correlation times of the quantities of interest become difficult to estimate due the sparsity of the data. We therefore expect the quality of the ad-hoc statistical nudging method to decrease.

\begin{figure}[h!]
    \centering
    \includegraphics[width=0.98\linewidth]{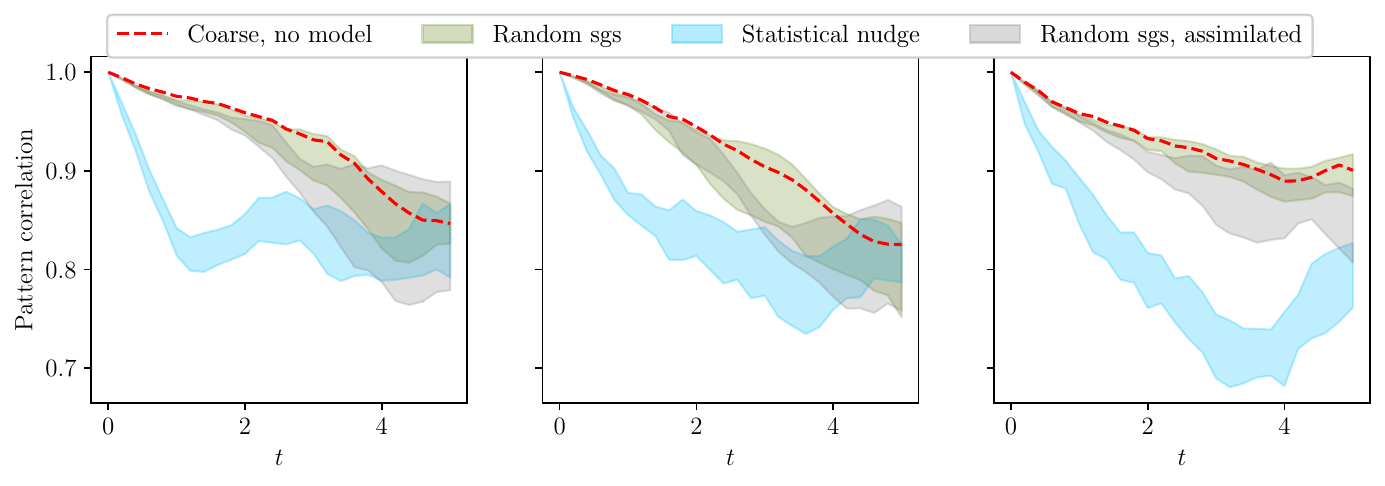}
    \caption{Pattern correlation between the prediction and the reference solution, using few data to calibrate the model. Three different initial conditions are considered. Each ensemble consists of 10 members; each band is colored between the maximal and minimal measured values.}
    \label{fig:few_pcorr}
\end{figure}

The pattern correlations for short lead times are given in figure \ref{fig:few_pcorr}. No significant change in prediction quality is observed, compared to the predictions based on plenty data in figure \ref{fig:plenty_pcorr}. This suggests that the prediction of the instantaneous solution at short lead times is robust under changes in the available data and instead relies on the accuracy of the initial condition.

\begin{figure}[h!]
    \centering
    \includegraphics[width=0.98\linewidth]{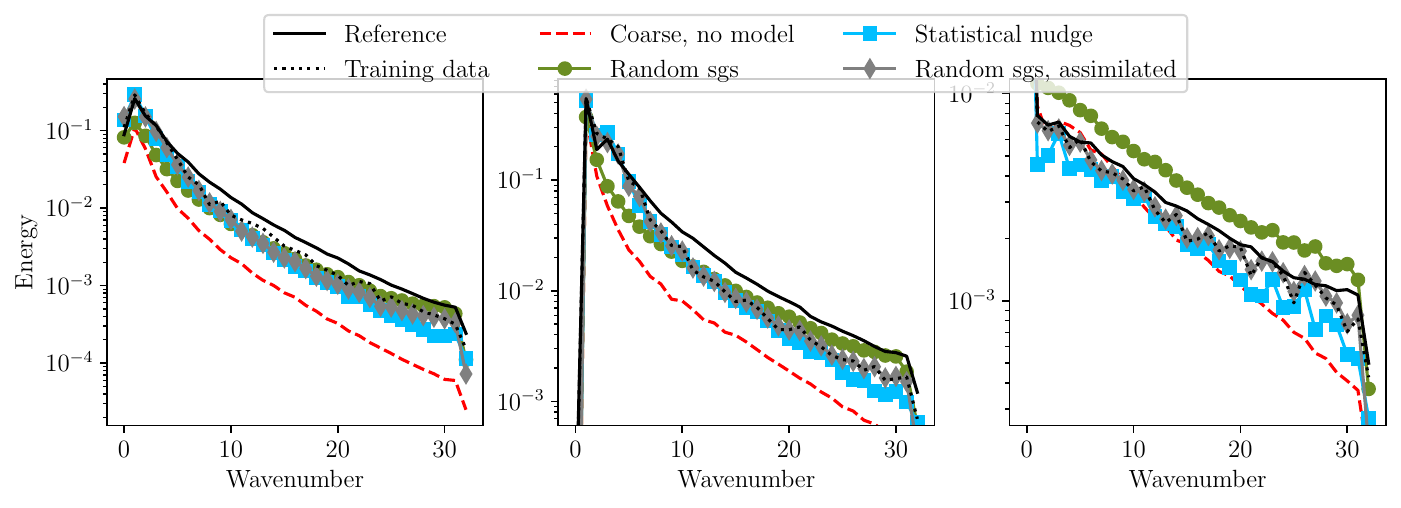}
    \caption{Time-averaged energy spectra measured along a horizontal cross-section of the domain for the horizontal velocity (left), vertical velocity (middle) and temperature (right). The cross-sections are take in the core of the domain at $y=5.5\times 10^{-1}$ for the horizontal velocity and $y=5.0\times 10^{-1}$ for the vertical velocity and the temperature.}
    \label{fig:few_spectra}
\end{figure}

\begin{figure}[h!]
    \centering
    \includegraphics[width=0.98\linewidth]{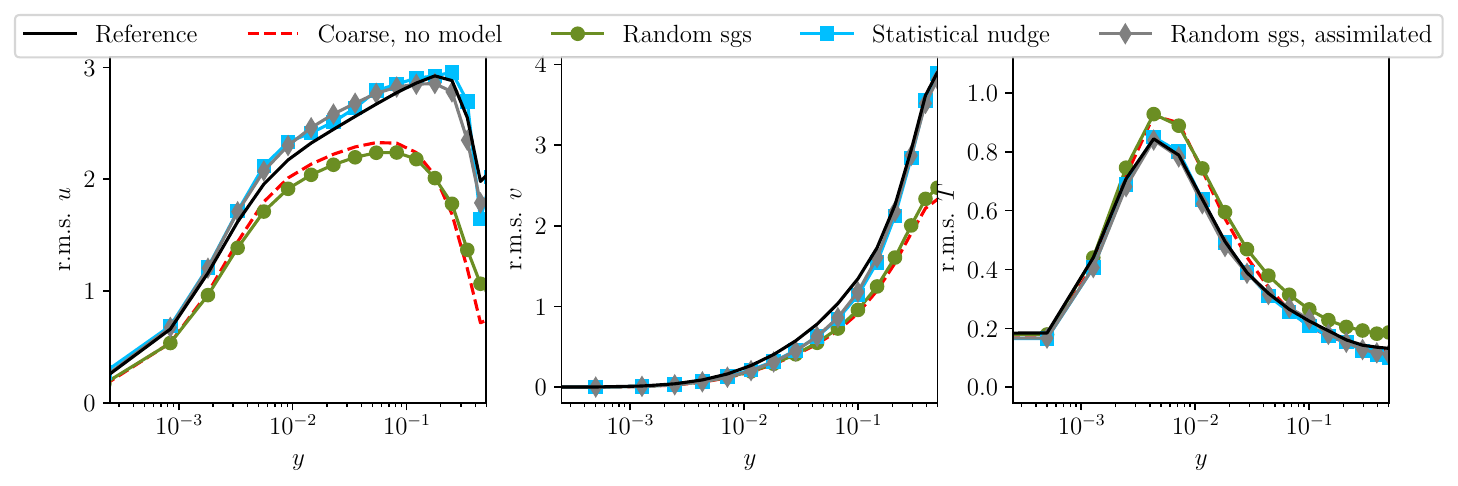}
    \caption{Average root mean square (r.m.s.) of the horizontal velocity (left), vertical velocity (middle) and temperature (right), measured along horizontal cross-sections of the domain and shown as functions of the wall-normal distance.}
    \label{fig:few_rms}
\end{figure}

\begin{figure}[h!]
    \centering
    \includegraphics[width=0.98\linewidth]{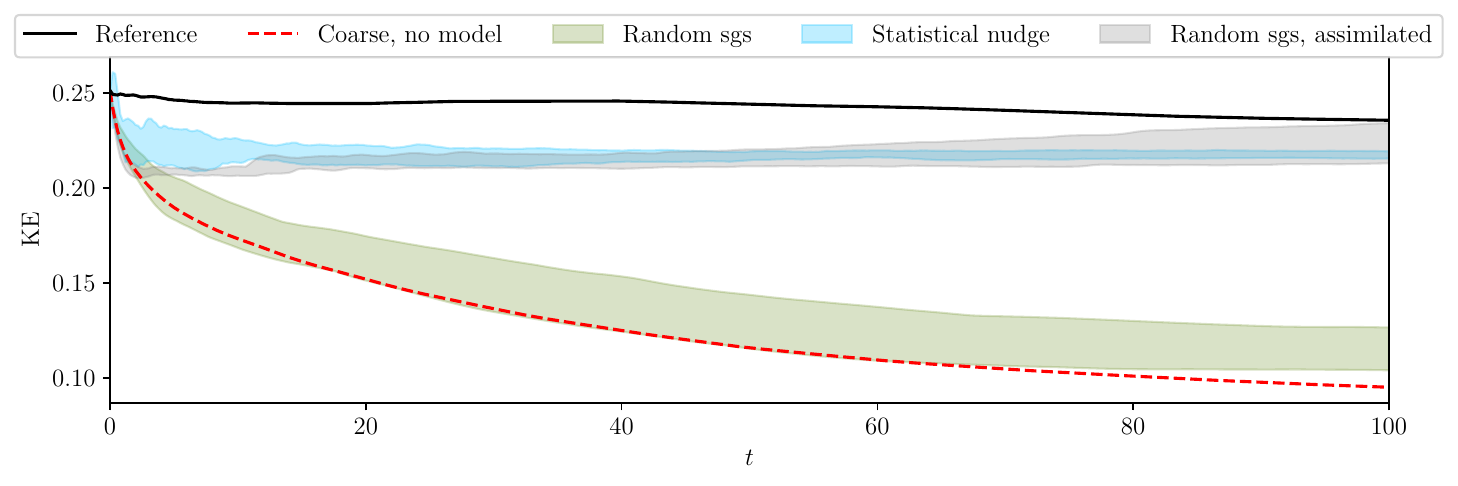}
    \caption{Rolling mean of the kinetic energy (KE) number over time.}
    \label{fig:few_kerm}
\end{figure}

\begin{figure}[h!]
    \centering
    \includegraphics[width=0.98\linewidth]{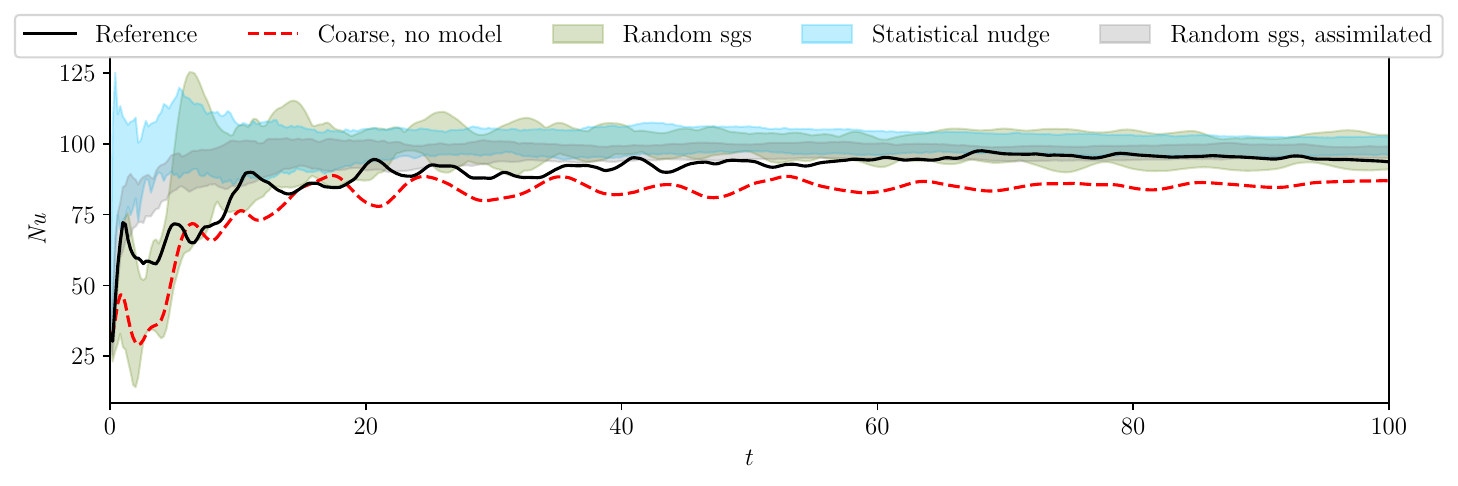}
    \caption{Rolling mean of the Nusselt number ($\mathit{Nu}$) over time.}
    \label{fig:few_nurm}
\end{figure}

The energy spectra in the core of the domain and the average r.m.s. values for the prognostic variables are depicted in figures \ref{fig:few_spectra} and \ref{fig:few_rms}, respectively. Good agreement is observed at the largest scales of motion despite the small amount of data, and the methods that employ a statistical correction adequately reproduce the energy in these scales. The effects of using limited data become apparent in the smaller scales of motion, particularly visible in the spectrum of the temperature. The average measured energy in these scales is not converged and hence deviate from the reference. The ensemble with assimilated statistics closely follows the training data, for which an improvement over the no-model result is still evident. The average r.m.s. values do not deteriorate using the small data set and show good agreement with the reference. The rolling averages of the kinetic energy and the Nusselt number, respectively shown in figures \ref{fig:few_kerm} and \ref{fig:few_nurm}, display the same qualitative behavior is observed as when using plenty data. Overall, no distinct loss of predictive quality is found using few data to calibrate the model when compared to using plenty data.


\subsection{Dependence on ensemble size}\label{subsec:ensemblesize}
\begin{figure}[h!]
    \centering
    \includegraphics[width=0.98\linewidth]{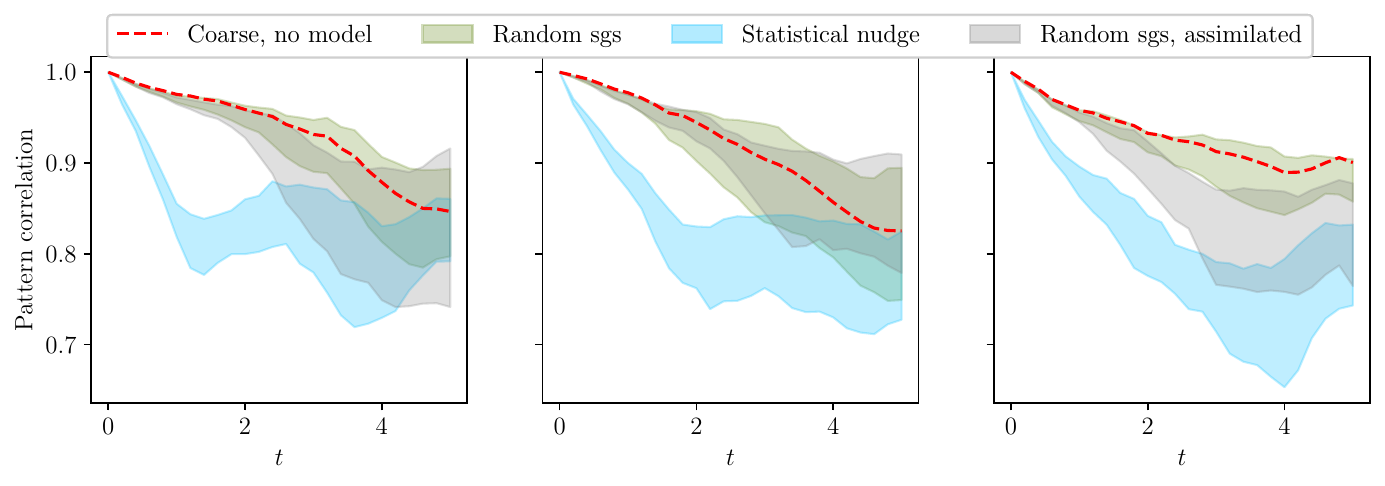}
    \caption{Pattern correlation between the prediction and the reference solution, using plenty data to calibrate the model. Each ensemble consists of 50 members; each band is colored between the maximal and minimal measured values.}
    \label{fig:largeensemble_pcorr}
\end{figure}

The EnKF results in the optimal linear estimator in the limit of large ensemble size \cite{law2016deterministic, schillings2017analysis}. Carrying out an ensemble simulation with many ensemble members might quickly become prohibitively expensive, even when using a computationally cheap low-fidelity solver. In the EnKF, the evolution of the individual ensemble members is coupled through the analysis step which uses information of the entire ensemble, and a change in system dynamics may thus be observed when changing the number of ensemble members. We therefore repeat the short-time numerical simulations presented in figure \ref{fig:plenty_pcorr} in Section \ref{subsec:plentydata} using 50 ensemble members instead of 10 to establish that a modest ensemble size does not adversely affect the model performance. The obtained pattern correlations in figure \ref{fig:largeensemble_pcorr} show no qualitative change with respect to the earlier presented results. This suggests that the currently adopted approach already provides robust forecasts at small ensemble size.

\section{Conclusions}\label{sec:conclusions}
In this paper, we have proposed a method for deriving probabilistic data-driven turbulence closure models suitable for coarsened steady-state turbulence. 
Based on ideal large-eddy simulation, a combination of stochastic large-eddy similation and data assimilation methods is suggested, requiring both \textit{a posteriori} collected data of the employed numerical solver and statistical data of the high-fidelity solution. 
Thus, the method exploits knowledge of the local integration error and the desired flow statistics.

The model was demonstrated using a non-intrusive implementation applied to two-dimensional \RB convection at Rayleigh number $\mathit{Ra}=10^{10}$. 
Stochastic perturbations based on sub-grid scale data were used in conjunction with a simplified ensemble Kalman filter to steer coarse numerical predictions towards desired statistics known from a precursor simulation. 
Here, we focused on horizontal energy spectra and average heat flux, which were found to be accurately reproduced with the proposed model. 
The model showed robust results for a modest ensemble size. 
No considerable deterioration of the predictions was observed using as few as 20 high-fidelity snapshots to determine the model parameters.

The presented modelling framework is general and can be applied to different fluid dynamical models. 
Nonetheless, many modelling choices have to be made in the actual implementation. Exploring different modelling choices is a challenge for future work. 
For example, more elaborate methods of \textit{a posteriori} measurements have been used in recent studies (see \cite{sanderse2024scientific} and references therein), which can aid the quality of the ensemble LES prediction.
Additionally, the current adopted implementation disregarded all covariances between the chosen quantities of interest. 
A promising approach to include these covariances without substantially increasing the computational cost associated to the model, is to significantly reduce the number of quantities of interest \cite{hoekstra2024reduced}. 
Adopting a smaller number of key statistics additionally makes it more tractable to compare different data assimilation schemes within the modelling framework.

\paragraph*{Software and data availability}
The data and adopted implementations that support the findings of this study are publicly available in Zenodo at \url{http://doi.org/10.5281/zenodo.13353273}.
\paragraph*{Acknowledgements} This work was supported by the Swedish Research Council (VR) through grant no. 2022-03453.
\paragraph*{Declaration of interests.} The authors report no conflict of interest.

\printbibliography

\appendix
\section{Error estimates for the adopted model}
\label{app:error_estimates}
Below, we provide error estimates for the model described in Section \ref{subsec:rb_model}.

\paragraph*{Notation and definitions.} The following estimate concerns a single quantity of interest $G$. Since each quantity of interest is updated independently of the other quantities, the analysis can be applied to each quantity of interest separately.

Let $G^n$ and $G^{n+1}$ be the quantity of interest at time instances $t^n$ and $t^{n+1}=t^n+\Delta t$, respectively. 
The evolution of $G$ is denoted by $\dot{G}=L_G(G)$. 
We consider an ensemble of $N$ members, where each member is denoted by a subscript $i, i=1,\ldots, N$. 
We assume the deterministic operator $L_G$ is Lipschitz continuous with Lipschitz constant $C_G$. 
Thus, we can use \begin{equation}
    G^n_i - \Delta t C_G \leq G^{n+1}_i \leq G^n_i + \Delta t C_G, \label{eq:lipschitz_ineq}
\end{equation}
for each ensemble member.

The stochastic perturbation is given by $\Delta t\mup + \sigp\rp^n$, where $\rp^n\sim \mathcal{N}(0, 1)$ and $\mup$ is the mean measured deviation after an integration step, specified as a model parameter. 
The observation at time instance $n$ is denoted by $G^n_{i,\mathrm{obs}} = \muo + \sigo\ro^n$, where $\ro^n\sim\mathcal{N}(0,1)$.
The values of $\rp^n$ and $\ro^n$ are i.i.d. for each $i$ and $n$.

The predicted value of ensemble member $i$ is given by $G_{i,f}$, \begin{equation}
     G_{i,f} = \int_{t^n}^{t^{n+1}}\! L_G(G^n_i)\,\td t + \Delta t \mup + \sigp\rp^n
\end{equation}

We denote the gain by $K$ and observe that this value is the same for each ensemble member. 
The ensemble and the observations have nonzero variance, from which it follows that $0<K<1$. We expand \begin{equation}
    \begin{split}
        G_i^{n+1} &= G_{i,f} + K (G^{n+1}_{i,\mathrm{obs}} - G_{i,f}) \\
        & =(1-K)\left(\int_{t^n}^{t^{n+1}}\! L(G^n_i)\,\td t + \Delta t \mup + \sigp\rp^n\right) + K G^{n+1}_{i,\mathrm{obs}},
    \end{split}
\end{equation}
which is used to bound the errors of individual ensemble members and corresponding observations, as well as the ensemble mean and the mean observation.

\paragraph{Error between individual ensemble members and measurements.}
We define the errors as the distance between the value of the quantity of interest at a time instance and the corresponding observation, that is, $E^{n+1}_i = G^{n+1}_i - d^{n+1}_i $ and $E^n_i = G^n_i - d^n_i$. We find \begin{equation}
    E^{n+1}_i = G^{n+1}_i - d^{n+1}_i = (1-K)\left(\int_{t^n}^{t^{n+1}}\! L(G^n_i)\,\td t + \Delta t \mup + \sigp\rp^n - G^{n+1}_{i,\mathrm{obs}} \right).
\end{equation}
Note that $G^{n+1}_{i,\mathrm{obs}} - G^n_{i,\mathrm{obs}} = \sigo\left(\ro^n - \ro^n\right)$ and therefore, using \eqref{eq:lipschitz_ineq}, we obtain \begin{equation}
    \begin{split}
        \left| E_i^{n+1} \right| &< \left| G_i^n + \Delta t C_G + \Delta t \mup + \sigp \rp^n - G^{n+1}_{i,\mathrm{obs}} \right| \\
        & \leq \left| G_i^n + \Delta t C_G + \Delta t \mup + \sigp \rp^n - G^n_{i,\mathrm{obs}} -\sigo\left(\ro^{n+1} - \ro^n\right) \right| \\
        &= \left| E_i^n + \Delta t C_G + \Delta t \mup + \sigp \rp^n -\sigo\left(\ro^{n+1} - \ro^n\right) \right| \\
        & \leq \left|E_i^n \right| + \Delta t \left|C_G + \mup \right| + \sigp |\rp^n| + \sigo |\ro^{n+1}-\ro^n|.
    \end{split}
\end{equation}

\paragraph{Error between ensemble mean and mean observation.}
Recall the ensemble mean of a quantity $f$ with ensemble members $f_i, ~i=1,\ldots,N$, \begin{equation}
    \bar{f} = \frac{1}{N}\sum_{i=1}^N f_i.
\end{equation}
We now study the development of distance between the ensemble mean of the quantity of interest and the measured mean value, i.e., $\bar E^{n+1}=|\bar G^{n+1}-\muo|$. We find that \begin{equation}
    \begin{split}
        E^{n+1}&=|\bar G^{n+1}-\muo| \\
        &= \left|\frac{1}{N}\sum_{i=1}^N (1-K)\left[\int_{t^n}^{t^{n+1}}\! L(G_i^n)\,\td t + \Delta t\mup + \sigp\rp^n - \muo \right] + K\sigo\ro^{n+1} \right| \\
        &\leq (1-K) \left|\frac{1}{N}\sum_{i=1}^N G_i^n - \muo \right| + (1-K) \left|\frac{1}{N}\sum_{i=1}^N \Delta t C_G + \Delta t \mup +\sigp\rp^n \right| + K\sigo\left|\frac{1}{N} \sum_{i=1}^N \ro^{n+1} \right| \\
        &< \left|\bar G^n - \muo \right| + \Delta t\left(C_G + \mup \right) + \sigp\left|\overline{\rp^n} \right| + \sigo \left|\overline{\ro^{n+1}} \right| \\
        &=E^n+ \Delta t\left(C_G + \mup \right) + \sigp\left|\overline{\rp^n} \right| + \sigo \left|\overline{\ro^{n+1}} \right|
    \end{split}
\end{equation}

Since $\rp^n$ and $\ro^n$ are i.i.d. with mean zero for each $i, n$, we observe that $\overline{\rp^{n+1}}$ and $\overline{\ro^{n+1}}$ approach zero as the ensemble size $N$ grows.
Thus, in the limit of infinite ensemble size, we obtain \begin{equation}
    E^{n+1} < E^n + \Delta t (C_G + \mup).
    \label{eq:mean_reversion}
\end{equation}
Equation \eqref{eq:mean_reversion} shows that the ensemble mean of $G$ reverts to $\muo$ when the ensemble size is sufficiently large and the time step is sufficiently small.

\end{document}